\documentclass[twocolumn,twocolappendix]{aastex631} 

\usepackage{graphicx}	
\usepackage{amsmath}	
\usepackage{amssymb}	
\usepackage{xspace}
\let\tablenum\relax
\usepackage{siunitx}
\usepackage{comment}
\usepackage{textcomp}
\usepackage{bm}
\usepackage{xcolor}

\newcommand{\gDor}{$\gamma$~Dor\xspace}
\newcommand{\gmodes}{g~modes\xspace}
\newcommand{\Msun}{\,M$_{\odot}$\xspace}
\newcommand{\dMsun}{-M$_{\odot}$\xspace}

\newcommand{\grad}{$g_{{\rm rad}, i}$\xspace}

\newcommand{\xci}{$X_{\rm c}/X_{\rm ini}$\xspace}
\newcommand{\zini}{$Z_{\rm ini}$\xspace}
\newcommand{\fov}{$f_{\rm ov}$\xspace}
\newcommand{\aov}{$\alpha_{\rm ov}$\xspace}

\newcommand{\kapr}{$\kappa_{\rm R}$\xspace}
\newcommand{\kaprt}{$\kappa_{\rm R}^{\rm (t)}$\xspace}
\newcommand{\temt}{$T^{\rm (t)}$\xspace}
\newcommand{\rhot}{$\rho^{\rm (t)}$\xspace}
\newcommand{\temr}{$T(r)$\xspace}
\newcommand{\rhor}{$\rho(r)$\xspace}
\newcommand{\gammat}{$\gamma^{\rm (t)}$\xspace}
\newcommand{\kaprp}{$\kappa_{\rm R}^{\rm (p)}$\xspace}

\newcommand{\mesa}{\texttt{MESA}\xspace}
\newcommand{\gyre}{\texttt{GYRE}\xspace}
\newcommand{\ester}{\texttt{ESTER}\xspace}
\newcommand{\kico}{KIC\,11294808\xspace}

\newcommand{\edits}[1]{#1}
\newcommand{\redits}[1]{#1}

\shorttitle{Shear mixing and radiative levitation}
\shortauthors{Mombarg et al.}
\graphicspath{{./}{figures/}}
\received{2021 Oct 20}
\revised{2021 Nov 19}
\accepted{2021 Nov 24}
\submitjournal{ApJ}

\begin{document}

\title{Predictions for gravity-mode periods and surface abundances in intermediate-mass dwarfs from shear mixing and radiative levitation }

\correspondingauthor{Joey S. G. Mombarg}
\email{joey.mombarg@kuleuven.be}

\author[0000-0002-9901-3113]{Joey S. G. Mombarg}
\affiliation{Institute of Astronomy, KU Leuven, Celestijnenlaan 200D, Leuven, Belgium}
\affiliation{IRAP, Universit\'e de Toulouse, CNRS, UPS, CNES, 14 avenue \'Edouard Belin, F-31400 Toulouse, France}

\author[0000-0002-4442-5700]{Aaron Dotter}
\affiliation{Department of Physics and Astronomy, Dartmouth College, Hanover, NH 03755 USA}

\author[0000-0002-9395-6954]{Michel Rieutord}
\affiliation{IRAP, Universit\'e de Toulouse, CNRS, UPS, CNES, 14 avenue \'Edouard Belin, F-31400 Toulouse, France}

\author[0000-0001-9097-3655]{Mathias Michielsen}
\affiliation{Institute of Astronomy, KU Leuven, Celestijnenlaan 200D, Leuven, Belgium}

\author[0000-0003-2771-1745]{Timothy Van Reeth}
\affiliation{Institute of Astronomy, KU Leuven, Celestijnenlaan 200D, Leuven, Belgium}

\author[0000-0003-1822-7126]{Conny Aerts}
\affiliation{Institute of Astronomy, KU Leuven, Celestijnenlaan 200D, Leuven, Belgium}
\affiliation{Department of Astrophysics, IMAPP, Radboud University Nijmegen, PO Box 9010, 6500 GL Nijmegen, The Netherlands}
\affiliation{Max Planck Institute for Astronomy, Koenigstuhl 17, 69117 Heidelberg, Germany}


\begin{abstract}

The treatment of chemical mixing in the radiative envelopes of intermediate-mass stars has hardly been calibrated so far. Recent asteroseismic studies demonstrated that a constant diffusion coefficient in the radiative envelope is not able to explain the periods of trapped gravity modes in the oscillation spectra of $\gamma$ Doradus pulsators. We present a new generation of \mesa stellar models with two major improvements. \redits{First, we present} a new implementation for computing radiative accelerations and Rosseland mean opacities that requires significantly less CPU time. \redits{Second,} the inclusion of shear mixing based on rotation profiles computed with the 2D stellar structure code \ester \redits{~is considered}. We show predictions for the mode periods of these models \redits{covering stellar masses} from 1.4 to 3.0\Msun across the main sequence (MS), computed for different metallicities. The morphology of the chemical mixing profile resulting from shear mixing in combination with atomic diffusion and radiative levitation does allow for mode trapping, while the diffusion coefficient in the outer envelope is large ($>10^{6}\,{\rm cm^2\,s^{-1}}$). Furthermore, we make predictions for the evolution of surface abundances for which radiative accelerations can be computed. \edits{We find that the N/C and C/O abundance ratios correlate with stellar age. We predict that these correlations are observable with precisions $\lesssim 0.1$\,dex on these ratios, given that a precise age estimate can be made.}   
\end{abstract}



\section{Introduction} \label{sec:intro}
A complete and calibrated theory of chemical mixing inside stars remains an outstanding problem in stellar structure and evolution theory \citep[][]{Salaris2017}. The chemical mixing on a microscopic level is ascribed to atomic diffusion (including radiative levitation) and can be derived from first principles, whereas macroscopic mixing, \edits{caused by turbulent flows}, is often parameterized as the combined effect of convection, (convective) core boundary mixing (CBM), and radiative envelope mixing (REM), each with their own free parameters. Convective mixing can be described by Mixing Length Theory \citep[MLT;][]{BohmVitense1958}, where the free parameter $\alpha_{\rm MLT}$, setting the typical length scale, is often fixed to a value calibrated to the Sun \citep{Choi2018}. A commonly used parametrization for the CBM is based on convective penetration \citep{Zahn1991, AugustsonMathis2019}, where the fully mixed core is extended over a distance described by a dimensionless parameter \aov times the local pressure scale height. In this extended region, the temperature gradient is the adiabatic one. \redits{This form of CBM was found to give a more accurate description than diffusive exponential overshooting with a radiative temperature gradient from a sample of 26 Slowly Pulsating B-type (SPB) stars \citep{Pedersen2021}.} Beyond the CBM zone, REM takes over which can be induced by several mechanisms, such as internal gravity waves \citep{RogersMcElwaine2017}, meridional flows \citep{Meynet2002}, or shear instabilities due to differential rotation \citep{maeder2009}. \newline

The advent of asteroseismology has made it possible to probe the internal physics of pulsating stars through their oscillation spectrum \citep{Aerts2010}. For non-rotating, non-magnetic stars the period difference between gravity (g) modes with equal spherical degree, $\ell$, azimuthal order, $m$, and consecutive radial order, $n$, is constant in the asymptotic regime where $n \gg \ell$ \citep{Tassoul1980}. Yet, departures from a constant period spacing are induced by the changes in the local composition \citep{Miglio2008}, by rotation \citep{Bouabid2013}, and by the presence of a magnetic field \citep{Prat2019,VanBeeck2020}. The study of a star's period spacing pattern (i.e. mode period $P_{n}$ versus the period spacing $\Delta P_{n} = P_{n+1} - P_{n}$), or equivalently the precise mode periods themselves, allows one to probe the deep internal stellar structure \citep[see][for a review]{Aerts2021}.

Different prescriptions for REM have been tested on a sample of SPB stars in the work by \cite{Pedersen2021}, comparing the observed periods of g-mode oscillations with those predicted by models. Their work shows a \edits{REM profile} based on vertical shear is preferred for the majority of the sample containing 26 stars. Similarly, \cite{Mombarg2021} measured the value of \fov, the parameter defining the extent of the \edits{CBM} region \citep{Freytag1996}, for a sample of 37 $\gamma$~Doradus (\gDor) stars. In their work, a constant value $D_{\rm macro} = 1\,{\rm cm^2\,s^{-1}}$ for the macroscopic mixing in the radiative envelope was chosen. Although for a large part of the sample this simple prescription seems sufficient, there are also several stars which show signatures of {\it mode trapping}, such that modes are confined in a narrow cavity due to \redits{wave} reflection caused by a gradient in the local mean molecular weight. The clearest example of a \gDor star with trapped modes is \kico, which shows pronounced `dips' in the period spacing pattern \citep{VanReeth2015-spec} that could not be explained with the physics used in the models of \cite{Mombarg2021}. \edits{These authors find that similar amplitudes of the dips, compared to those observed in \kico, are only reproduced at a level of $D_{\rm macro} = 0.05\,{\rm cm^2\,s^{-1}}$ for the macroscopic mixing.}
Yet, such low diffusion coefficients throughout the radiative envelope is four to six orders of magnitude lower than typical values assumed for \edits{macroscopic mixing induced by} turbulent diffusion in stellar models \citep{Miglio2008,Christophe2018,Ouazzani2019}. Furthermore, due to gravitational settling, \kico should have a low surface metallicity (if radiative levitation is relatively weak), which is not supported by spectroscopic observations \citep{Gebruers2021}. \newline

Moreover, the work by \cite{Mombarg2020} demonstrated the inclusion of atomic diffusion and radiative levitation has large effects on the resulting predicted periods of the \gmodes. \redits{Similar results were obtained for p~modes by \cite{Deal2020}.} Yet, due to large amounts of computation time required for calculations related to radiative levitation and the Rosseland mean opacity, the inclusion of radiative levitation in the modeling of \gmodes in dwarfs has so far been limited to two slowly rotating \gDor stars \citep{Mombarg2020}.  \newline
 The gravitational settling of helium near the core boundary reduces the local chemical gradient, and therefore inhibits mode trapping \citep{Theado2009}. Since it is well-known that atomic diffusion occurs in stars \citep{Michaud2015}, this gives us extra constraints on the macroscopic mixing as the effect of atomic diffusion needs to be counteracted in some parts of the stars when mode trapping is observed. Similarly, a form of turbulent diffusion that is scaled with the local atomic diffusion coefficient of helium was introduced by \cite{Richer2000} to counteract the gravitational settling in AmFm stars, in order to reconcile with observed surface abundances. 
 
 \edits{In this work, we demonstrate that (macroscopic) shear mixing in combination with microscopic mixing \redits{due to atomic diffusion (including radiative levitation)} can explain the observed g-mode periods of \gDor stars, and we investigate the implications \redits{of including these two mixing phenomena} for the evolution of the surface abundances.  }
 In Section~\ref{sec:radlev}, we present an improved method to compute radiative accelerations, and the Rosseland mean opacity from monochromatic opacity tables, reducing the computation time drastically by more than a factor 4. In Section~\ref{sec:shear}, we discuss a new implementation of macroscopic mixing induced by vertical shear, and combine this with microscopic mixing to study when mode trapping can occur across the \gDor mass regime for different ages and metallicities (Section~\ref{sec:psp}). Moreover, in Section~\ref{sec:abun}, we show the predicted evolution of surface abundances, \redits{discuss the numerical sensitivity in Section~\ref{sec:numerical}}, and finally conclude in Section~\ref{sec:discussion}.

\section{Improved routines for radiative levitation} \label{sec:radlev}
\edits{We start by discussing the numerical implementation to compute radiative accelerations (i.e. the acceleration of elements induced by the process of radiative levitation}), and the improvements we have made to reduce the computation time.
The change in the local mass fraction of a chemical species $i$ over time is described by
\begin{equation}
\begin{split}
    & \frac{\partial X_i}{\partial t} =  \left( \frac{{\rm d}X_i}{{\rm d}t}  \right)_{\rm nuc} \\ &+  \frac{1}{\rho r^2}\frac{\partial}{\partial r}\left( \rho r^2 \left[D_{\rm macro}(t,r)\frac{\partial X_i}{\partial r} - v_{{\rm diff}, i}(t,r)\right] \right), \label{eq:nuc_syn2}
\end{split}    
\end{equation}
where the first term on the right-hand side describes the change in composition due to nuclear reactions, and the second term the change induced by both macroscopic and microscopic mixing, respectively. Here, \redits{$\rho$ is the local density, $r$ is the distance from the stellar center}, $D_{\rm macro}$ is the diffusion coefficient for macroscopic \edits{turbulent} mixing, and $v_{{\rm diff},i}$ is the diffusion velocity of species $i$.

We make use of the open-source stellar structure and evolution code \mesa, r11701 \citep[][]{Paxton2011, Paxton2013, Paxton2015, Paxton2018, Paxton2019}, where the default routines for computing radiative accelerations are based on the work by \cite{Hu2011}. 
The diffusion velocity of a species $i$ is determined in \mesa by solving the Burgers' equations \citep{Burgers1969}. We refer to \cite{Mombarg2020} and references therein for details \edits{on the numerical implementation}. The effectiveness of radiative acceleration enters Burgers' equations via the local acceleration induced by momentum absorption of photons generated in the stellar core \citep{Hu2011},
\begin{equation}
    g_{{\rm rad},i}(r) = \frac{\mu \kappa_{\rm R}}{\mu_i c} \mathcal{F}(r) \gamma_i(r),
\end{equation}
where $\mu_i$ and $\mu$ are the molecular weights of species $i$ and the mean molecular weight, respectively, $\kappa_{\rm R}$ is the Rosseland mean opacity, $\mathcal{F}(r)$ is the local radiative flux, $c$ is the speed of light, and 
\begin{equation} \label{eq:gamma}
    \gamma_i(r) = \int_0^\infty \frac{\sigma_i(u)[1 - e^{-u}] - a_i(u)}{\sum_k f_k(r) \sigma_k(u)}{\rm d}u,
\end{equation}
\edits{where $u=h\nu/k_{\rm B}T$ (with $h$ the Planck constant and $k_{\rm B}$ the Boltzmann constant).}
This integral depends on the local mixture which is given by the fractional abundances $f_k$, with $\sum_k f_k = 1$, and its computation requires the monochromatic cross sections $\sigma_i$, and correction terms $a_i$. \mesa relies on the tables provided by the OP project \citep{Seaton2005} for these quantities. These OP monochromatic tables provide data for H, He\footnote{For H and He, the correction terms $a_i$ are not provided.}, C, N, O, Ne, Na, Mg, Al, Si, S, Ar, Ca, Cr, Mn, Fe, and Ni, which are equally spaced in
\begin{equation} \label{eq:v_u}
    v(u^\prime) \equiv \frac{15}{4 \pi^4} \int_0^{u^\prime} \frac{u^4 e^{-u} {\rm d}u}{\left(1 - e^{-u}\right)^3},
\end{equation} 
where $u$ ranges from $10^{-3}$ to 20. 

As a result of atomic diffusion, the Rosseland mean opacity can no longer be evaluated for the same mixture in each cell of the model since,
\begin{equation}
    \kappa_{\rm R}(r) = \frac{1}{\mu} \left( \int \frac{1}{\sum_{k} f_k(r) \sigma_k(u)}  {\rm d}u \right)^{-1},
\end{equation}
where the factors $f_k(r)$ for the metals are no longer constant throughout the star.
Consistently computing $\kappa_{\rm R}$ requires a lot of additional computation time compared to when the fractional metal abundances are constant throughout the star.

\begin{table*}[]
    \centering
    \begin{tabular}{l|l}
        Quantity description & Dimension \\
        \hline
        Precomputed $\log \kappa_{\rm R}(\log T, \log \rho)$ for the mixture in the core. & 1648  \\
        Value of $(\log T(r), \log \rho(r))$ of the last time step for each cell. & $2 \times N_{\rm cell}$ \\
        Saved $\log \kappa_{\rm R}(\log T, \log \rho)$ grid for old mixture in cell. & $N_{\rm cell} \times 4 \times 4$ \\
        Precomputed $\log \gamma_i(\log T, \log \rho)$ for two averaged mixtures of the two zones in the radiative envelope. & $2 \times 1648$ \\

    \end{tabular}
    \caption{Summary of the quantities saved for the next time step in the evolution.}
    \label{tab:quan_save}
\end{table*}

As a first step to reduce the number of redundant computations, we pre-process the OP mono tables where we compute the cross section for momentum transfer to an atom (i.e. the numerator in Eq.~\ref{eq:gamma}). The data from the OP project are tabulated in temperature $T$ and electron number density $N_e$. The latter has to be converted to a mass density as per,
\begin{equation}
    \log \rho = \log N_e + \log \mu - \log \Xi - \log \mathcal{N}_{\rm A},
\end{equation}
where $\Xi = \sum_{k} f_k \Xi_k$ the average number of electrons per atom, and $\mathcal{N}_{\rm A}$ is Avogadro's number. 
For the temperature $T(r)$ and density $\rho(r)$ in each cell $\kappa_{\rm R}$ and $\gamma$ are computed by means of a bicubic interpolation, where we use the following scheme to select the 16 points on which the interpolation is done.
First, select the closest point in the table by minimizing

\begin{equation}
    \begin{split}
        & \Delta = \\ &\sqrt{\frac{(\log T^{\rm (t)} - \log T(r))^2}{0.0025} + \frac{(\log \rho^{\rm (t)} - \log \rho(r))^2}{0.25}},
    \end{split}    
\end{equation}
where the two constants are used to account for the difference in spacing of the tabulated \temt \redits{($\log T^{\rm (t)}$ given in steps of 0.05\,dex)} and \rhot \redits{($\log \rho^{\rm (t)}$ given in steps of 0.5\,dex)}. Subsequently, a grid of 4$\times$4 points is constructed, centered around $T(r)$ and $\rho(r)$. It occasionally happens that a cell has temperature and/or density close to the edge of the OP mono tables. In that case, the 4$\times$4 grid is moved one step at the time in temperature/density until a 4$\times$4 grid is found which encloses $T(r)$ and $\rho(r)$. For one location in the OP mono tables (relevant for intermediate-mass MS stars), the grid selection is hard-coded.

To further speed up the computation of \kapr, the quantity \kaprt is only recomputed if
\begin{equation} \label{eq:cond1}
    \max\limits_k \left(\frac{|f_k^{(c)} - f_k^{(p)}|}{f_k^{(p)}} \right) > 10^{-4},
\end{equation} 
where the superscript `(p)' indicates the mixture for which the grids have been precomputed, and `(c)' indicates the current mixture. The threshold value of $10^{-4}$ was found to be an optimal value by \cite{Hui-Bon-Hoa2021}, using a similar technique to speed up computations of the Rosseland mean opacity in the Toulouse-Geneva evolution code \citep{Hui-Bon-Hoa2008, Theado2012}. We choose to evaluate the mixture in the convective core, as this is the largest zone in the star where the mixture is homogeneous. After the first time step for which \kaprp is computed, the 4$\times$4 grids selected for each cell, and the corresponding \temr, \rhor and $f_k$ are saved for the next time step. For the following time steps, if Eq.~\ref{eq:cond1} is true, the following conditions are evaluated,
\begin{eqnarray}
 \max\limits_k \left(\frac{|f_k^{\rm (c)} - f_k^{\rm (o)}|}{f_k^{\rm (o)}} \right) &>& 10^{-4}, \\
 \left| \log T^{\rm (c)}(r) - \log T^{\rm (o)}(r)\right| &>& 10^{-2},\\
 \left| \log \rho^{\rm (c)}(r) - \log \rho^{\rm (o)}(r)\right| &>& 10^{-1},
\end{eqnarray}
where the superscript `(o)' indicates the values of the previous time step. If any of these three conditions is fulfilled for a specific cell, the grid points for interpolation are recomputed for the new local mixture, and stored for the next time step. Otherwise, the grid points of the previous time step are used for the interpolation of the opacity and its derivatives with respect to temperature and density. 

Similarly, for the radiative accelerations we precompute a grid containing \gammat for each temperature-density point in the OP mono data, using the initial mixture. Yet, as we only need to compute the radiative accelerations in the radiative envelope, a slightly different methodology is used, compared to the one used to compute the opacity. The radiative envelope is divided into two zones \edits{with equal number of cells} based on the number of cells for which \grad needs to be computed. In each of these two zones, we define a mixture, \edits{where $f_k = \sum_i f_{k,i}$ is the average value of the fractional abundance of element $k$ across all cells (with cell index $i$) in the zone, and renormalize afterwards such that $\sum_k f_k = 1$.
}

The value of $\gamma$ is then interpolated in temperature and density from one of the precomputed grids. A precomputed grid is recomputed for the current average mixture $f_k^{\rm (c)}$ if the condition in Eq.~\ref{eq:cond1} is true. \edits{To insure a smooth transition between these two zones, within 15 cells on each side of the boundary we blend the two values of $\log g_{{\rm rad},i}$ for the two average mixtures according to,
\begin{equation}
    \log g_{{\rm rad},i} = \beta_i \log g_{{\rm rad},i}^{(1)} + (1-\beta_i)\log g_{{\rm rad},i}^{(2)},
\end{equation}
where we vary $\beta_i$ linearly between 0 and 1 in this transition zone and the superscripts denote the two zones which each have an average mixture.}

\begin{figure*}
    \centering
    \includegraphics[width = 0.9\textwidth]{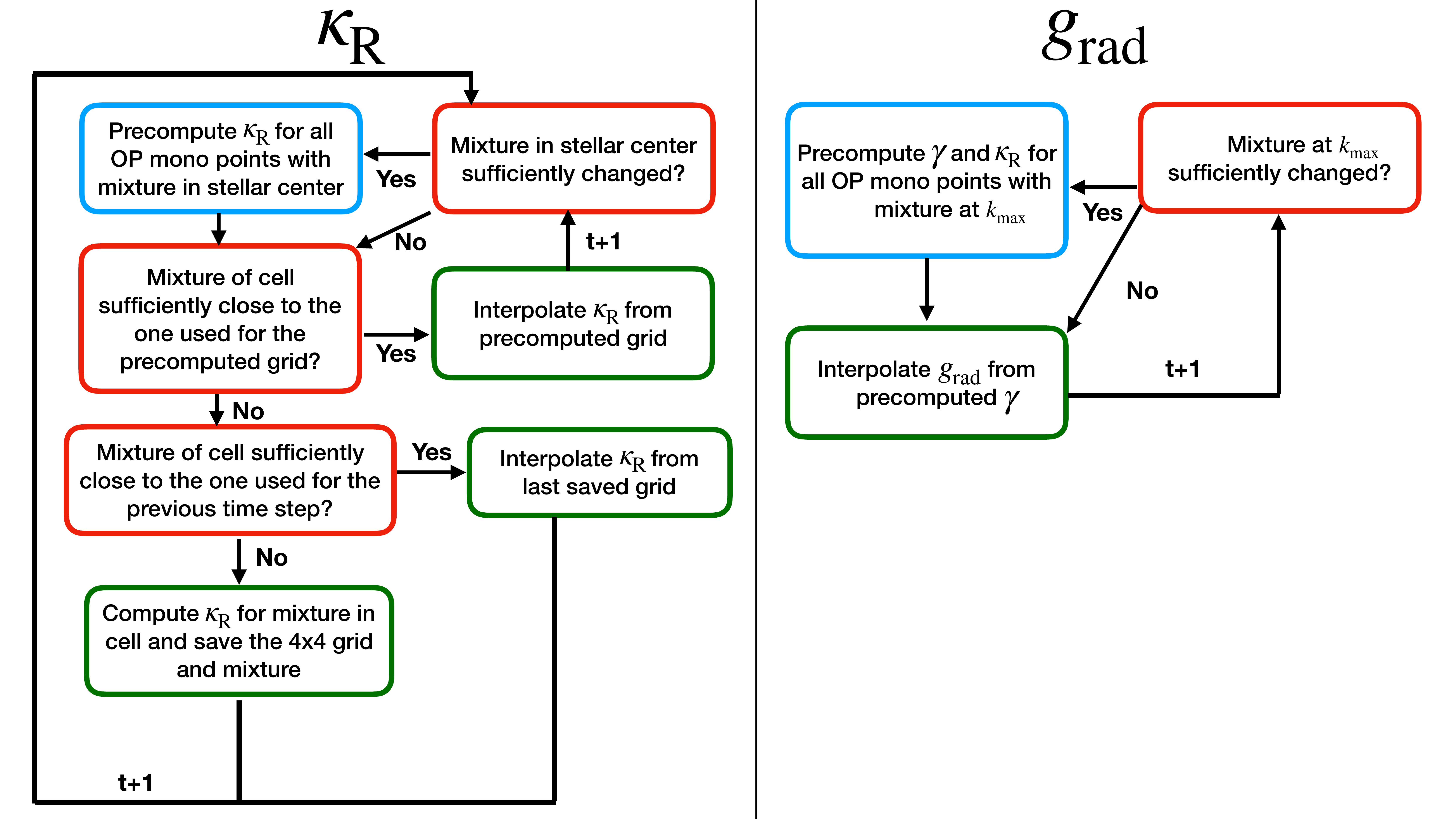}
    \caption{\edits{Flowchart showing the different steps in computing the Rosseland mean opacity (left side) and the radiative accelerations (right side), starting at the blue box. The labels $t+1$ indicate the next evolution time step.}}
    \label{fig:flowchart}
\end{figure*}

In summary, the quantities which are saved for the next time step are listed in Table~\ref{tab:quan_save} \edits{and a flowchart is shown in Fig.~\ref{fig:flowchart}}. Hence, by reusing previously made computations, and allowing a small error on the precision of the opacity and radiative accelerations, much CPU time is saved. To evolve a 1.7\dMsun model starting at the pre-MS to a core-hydrogen mass fraction of $X_{\rm c} = 0.005$, including radiative levitation for all elements for which OP mono data are available, takes 30\,min on 36 CPUs with the routine presented here. In comparison, this takes 130\,min with the current routines in \mesa. Additionally, models computed in this work experience almost no failures to converge on the pre-MS, which was until now, besides the long wall times, also a bottleneck \edits{for us} to compute grids of models with radiative levitation included. Fig.~\ref{fig:grad} in the appendix shows the difference in the predicted radiative accelerations and Rosseland mean opacity between the method presented in this paper and those of \cite{Hu2011}. Overall, the differences are small and the differences in the surface temperature and luminosity are well within typical precisions that can be achieved observationally. \clearpage

\section{Implementation of shear mixing in MESA} \label{sec:shear}
\begin{figure}
    \centering
    \includegraphics[width = 1.05\columnwidth]{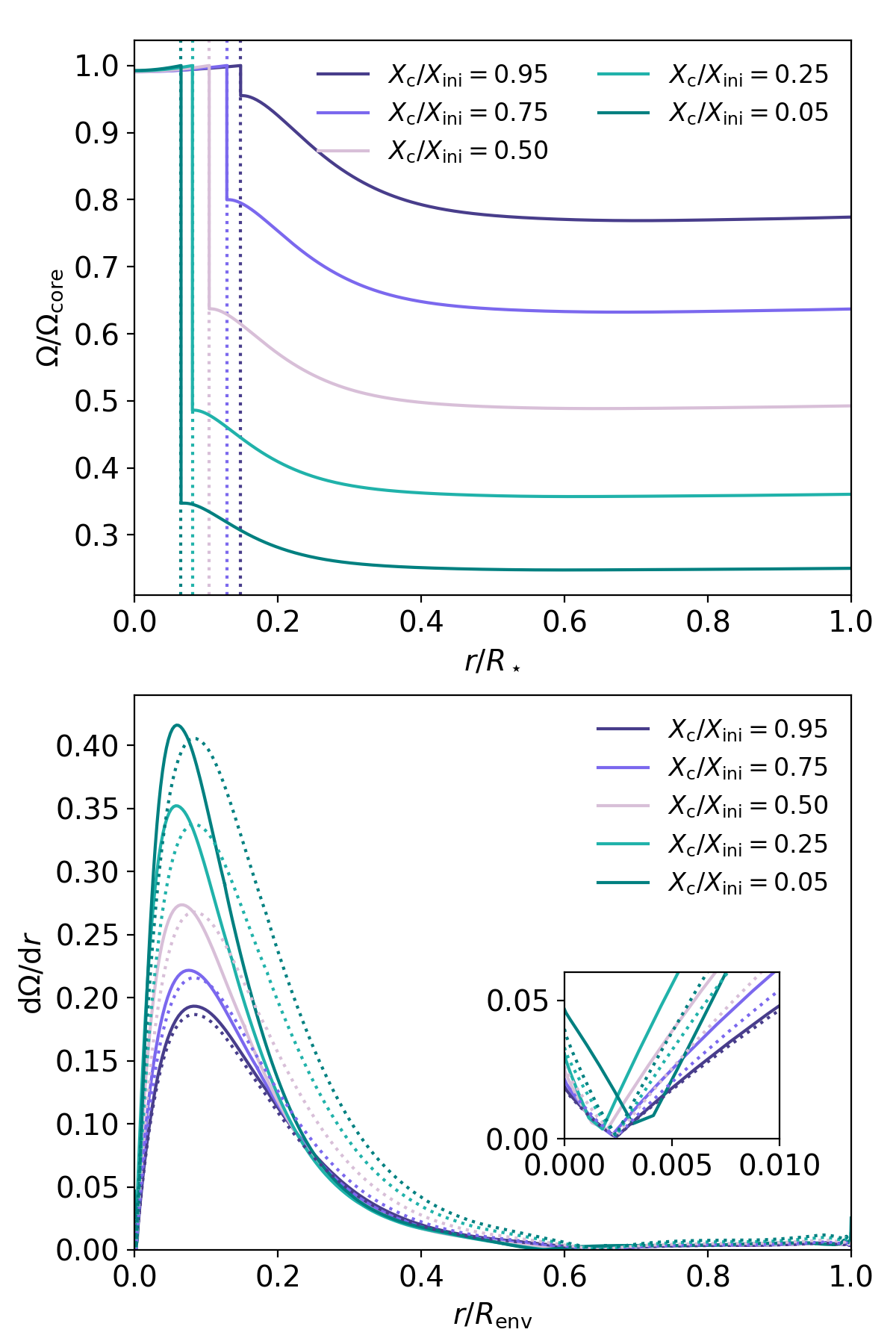}
    \caption{{\it Top panel:} Rotation profiles for a 3\dMsun \ester model with a rotation rate of $\Omega = 0.2\Omega_{\rm bk}$. The profiles are normalized by the rotation frequency at the core boundary (indicated by the dotted vertical lines). {\it Bottom panel:} The evolution of the shear profile (absolute value) for a 3\dMsun \ester model (solid lines) and the scaled approximations used for the mixing in the \mesa models (dotted lines). The inset shows a zoom-in close to the core boundary (i.e. where $r=0$). The radial derivative of the local rotation frequency shown here is dimensionless, as $\Omega$ is normalized by $\sqrt{P_{\rm c}/(\rho_{\rm c}R_\star^2})$, and $r$ is normalized by the size of the radiative envelope, $R_{\rm env}$, \edits{where $r = 0$ at the convective core boundary}.}
    \label{fig:Omega-r_Xc}
\end{figure}

In this work, we focus on macroscopic mixing from rotation-induced turbulence, henceforth referred to as shear mixing. It should be noted that this does not refer to the various forms of shear instabilities implemented in \mesa \citep[cf.][]{Heger2000}, which cause very spiky mixing profiles, and as such, are not suitable for g-mode asteroseismology \citep{Truyaert2016}.
We follow the formalism by \citet[][Eq. 2.14]{Zahn1992}, where we scale the local diffusive mixing coefficient as per
\begin{equation} \label{eq:Dmix}
    D_{\rm macro}(r,t) = \eta K \left( \frac{r}{N} \frac{{\rm d}\Omega}{{\rm d}r} \right)^2,
\end{equation}
where $\eta$ is a free parameter, $N$ is the (weighted) Brunt-V\"ais\"al\"a frequency, and $K$ the thermal diffusivity,
\begin{equation}
    K = \frac{16 \sigma_{\rm SB} T^3}{3 \kappa_{\rm R} \rho^2 C_{\rm P}},
\end{equation}
with $\sigma_{\rm SB}$ the Stefan-Boltzmann constant, and $C_{\rm P}$ the heat capacity.
The radial derivative of the rotation frequency, $\Omega$, is computed using the 2D \ester code \citep{Espinosa-Lara2013, Rieutord2016} that computes a \edits{steady} model at a given mass, $M_\star$, fraction of the initial core hydrogen mass fraction left, \xci, and rotation rate expressed as a fraction of the Keplerian critical rotation rate, $\Omega_{\rm bk}$. For stars with masses below $\sim 3$\Msun, the convergence of \ester becomes delicate. As such, we compute an \ester model for 3\Msun, $X_{\rm c}/X_{\rm ini} = 0.95$, and $\Omega_{\rm bk} = 0.2$ and apply simple scaling relations for $M_\star$ and $X_{\rm c}/X_{\rm ini}$ \edits{to scale the rotation profile from this model to the appropriate mass and age of the model we want to compute}. We take a rotation profile averaged over the co-latitude, starting from the boundary of the convective core, as shown in the top panel of Fig.~\ref{fig:Omega-r_Xc}. As can be seen in the bottom panel of Fig.~\ref{fig:Omega-r_Xc}, the global shear profile of $\frac{{\rm d}\Omega}{{\rm d}r}$ remains roughly the same along the evolution. Therefore, we choose to scale the normalized  $\left( \frac{{\rm d}\Omega}{{\rm d}r} \right)_{\rm norm}$ profile of the aforementioned \ester model as per,
\begin{eqnarray}
 \xi_{M_\star} &=& a_1 M_\star^2 + a_2 M_\star + a_3, \label{eq:xim}\\
 \xi_{X_{\rm c}/X_{\rm ini}} &=& a_4 (X_{\rm c}/X_{\rm ini})^2 + a_5 (X_{\rm c}/X_{\rm ini}) + a_6. \label{eq:xix}
\end{eqnarray}
That is, 
\begin{equation}
    \frac{{\rm d}\Omega}{{\rm d}r} (M_\star, X_{\rm c}/X_{\rm ini}) = \xi_{M_\star}\cdot \xi_{X_{\rm c}/X_{\rm ini}} \cdot \left(\frac{{\rm d}\Omega}{{\rm d}r}\right)_{\rm norm}.
\end{equation}

\begin{figure}
    \centering
    \includegraphics[width = 1.05\columnwidth]{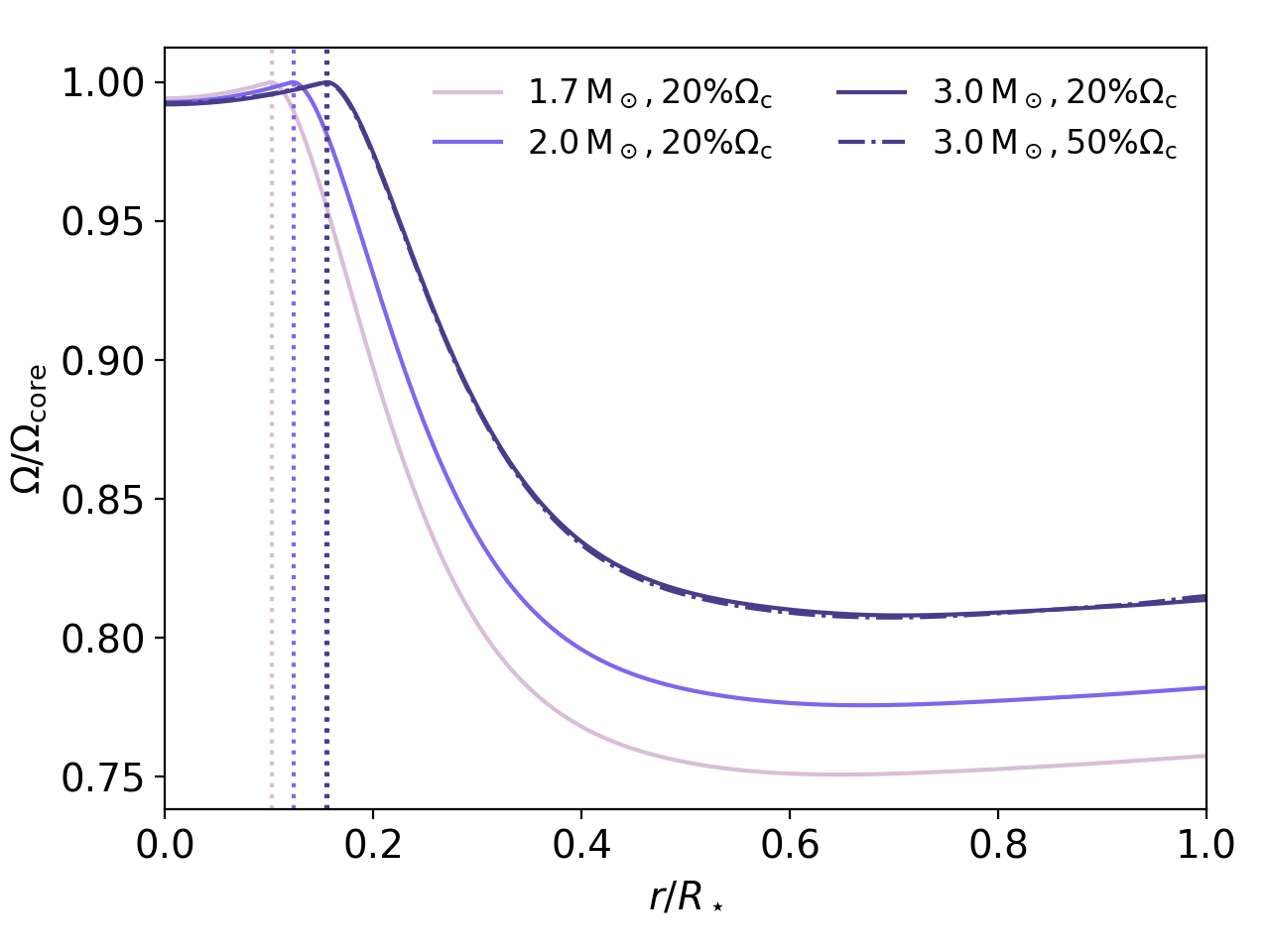}
    \caption{The variation of the normalized rotation profile with stellar mass for ZAMS models. For the 3.0\dMsun model, the rotation profile at $0.5\Omega_{\rm bk}$ is shown by the dashed-dotted line (almost the same as the profile for $0.2\Omega_{\rm bk}$). \edits{Note that in this plot $r \in [0, R_\star$].} }
    \label{fig:Omega-r_M}
\end{figure}

The values of the constants $a_i$ are listed in Table~\ref{tab:poly}. At each cell of the MESA model, the local value of $D_{\rm macro}$ is interpolated from the normalized $\frac{{\rm d}\Omega}{{\rm d}r}$ profile of the 3\Msun \ester model that is defined on an interval $[r_{\rm cc}, R_\star]$, that is, from the convective core boundary to the surface. In Fig.~\ref{fig:Omega-r_M}, we show rotation profiles for three different masses at the zero-age main sequence (ZAMS). In addition, the rotation profile for $0.5\Omega_{\rm bk}$ is also shown for the most massive model, demonstrating the morphology of the normalized rotation profile barely changes for different rotation rates.   \newline
\begin{table}[]
    \centering
    \begin{tabular}{c|r}
    Coefficient & Value \\
    \hline
        $a_1$ &   0.05159189\\
        $a_2$ &  -0.30399420\\
        $a_3$ &   0.63434263\\
        $a_4$ &   0.14382828\\
        $a_5$ &  -0.39459215\\        
        $a_6$ &   0.43750162\\        
    \end{tabular}
    \caption{Values of the coefficients shown in Equations~\ref{eq:xim} and \ref{eq:xix}.}
    \label{tab:poly}
\end{table}

\begin{figure*}
    \centering
    \includegraphics[width = 0.9\textwidth]{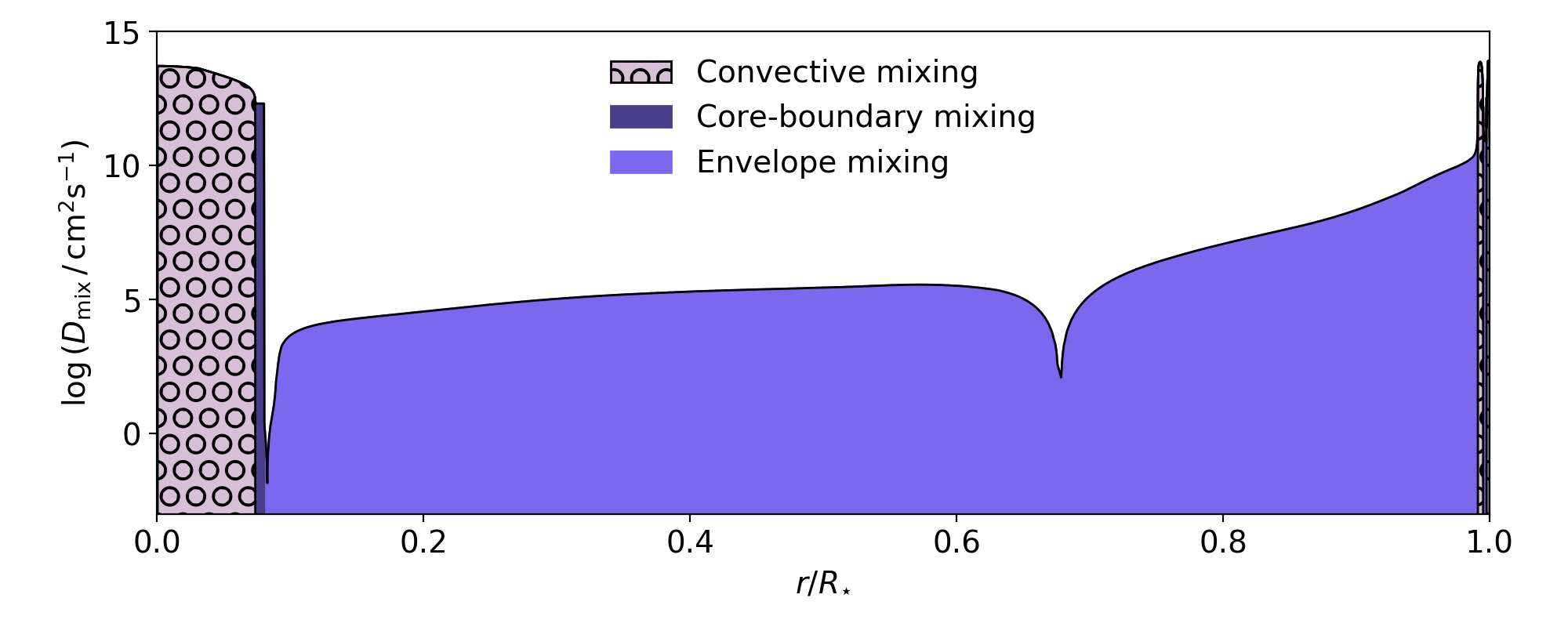}
    \caption{The diffusive mixing coefficient, $D_{\rm mix}$, throughout the star for a 1.7\dMsun model with \xci=0.45. The envelope mixing is computed according to Eq.~\ref{eq:Dmix} and atomic diffusion (including radiative levitation) have been taken into account.}
    \label{fig:Dmix}
\end{figure*}

The Brunt-V\"ais\"al\"a frequency can be decomposed into a part related to structure, and a part related to composition,
\begin{eqnarray}
 N_{\rm struc}^2 &\simeq& \frac{g\delta}{H_P}  \left(\nabla_{\rm ad} - \nabla\right), \\
 N_{\rm comp}^2 &\simeq&  \frac{g\phi}{H_P}  \nabla_\mu,
\end{eqnarray}
where $\nabla_{\rm ad}$, $\nabla$, and $\nabla_\mu$ are the adiabatic, temperature, and chemical gradient, respectively. Moreover, $g$ is gravitational acceleration, $H_P$ the local pressure scale height, $\delta = -(\partial \ln \rho / \partial \ln T)|_{P,\mu}$, and $\phi = (\partial \ln \rho / \partial \ln \mu)|_{P,T}$. \edits{ In the formalism of \cite{Mathis2004}, the mixing coefficient scales with $K + D_{\rm h}$, where $D_{\rm h}$ is the horizontal diffusion coefficient. Here, we do not compute $D_{\rm h}$, but instead approximate $K + D_{\rm h} = \eta K$, and construct a weighted Brunt-V\"ais\"al\"a frequency,}
\begin{equation}
    N^2 = N_{\rm struc}^2 + \frac{\eta}{\eta - 1}N_{\rm comp}^2.
\end{equation}
Hence, our implementation has one free parameter, $\eta$, to control the global scaling of the mixing profile. \redits{This weighting of the different components follows from the fact that while both components of the buoyancy force are weakened by horizontal diffusion, only the structure (thermal) component is additionally weakened through thermal diffusion \citep[][]{Talon1997}.} The point where the transition is made from CBM to REM is chosen at the smallest radius where $D_{\rm REM}(r) > D_{\rm ov}(r)$. An example of a profile for $D_{\rm macro}(r)$ is shown in Fig.~\ref{fig:Dmix}. The mixing profile based on vertical shear used by \cite{Pedersen2021} comes from a profile computed by \cite{Georgy2013}, relying on 1D stellar models. \cite{Pedersen2021} then scale this profile by a free parameter that sets the diffusion coefficient at the edge of the overshoot zone. In this work, we rely on 2D \ester models to compute the rotation profile, which are in addition scaled with the stellar mass and the hydrogen mass fraction in the core. We find similar mixing profiles compared to those derived by \cite{Mathis2004}, although the drop in the diffusion coefficient close to the overshoot zone (around $r/R_\star = 0.1$ in Fig.~\ref{fig:Dmix}) is much more pronounced in our models. 

\begin{figure}
    \centering
    \includegraphics[width = \columnwidth]{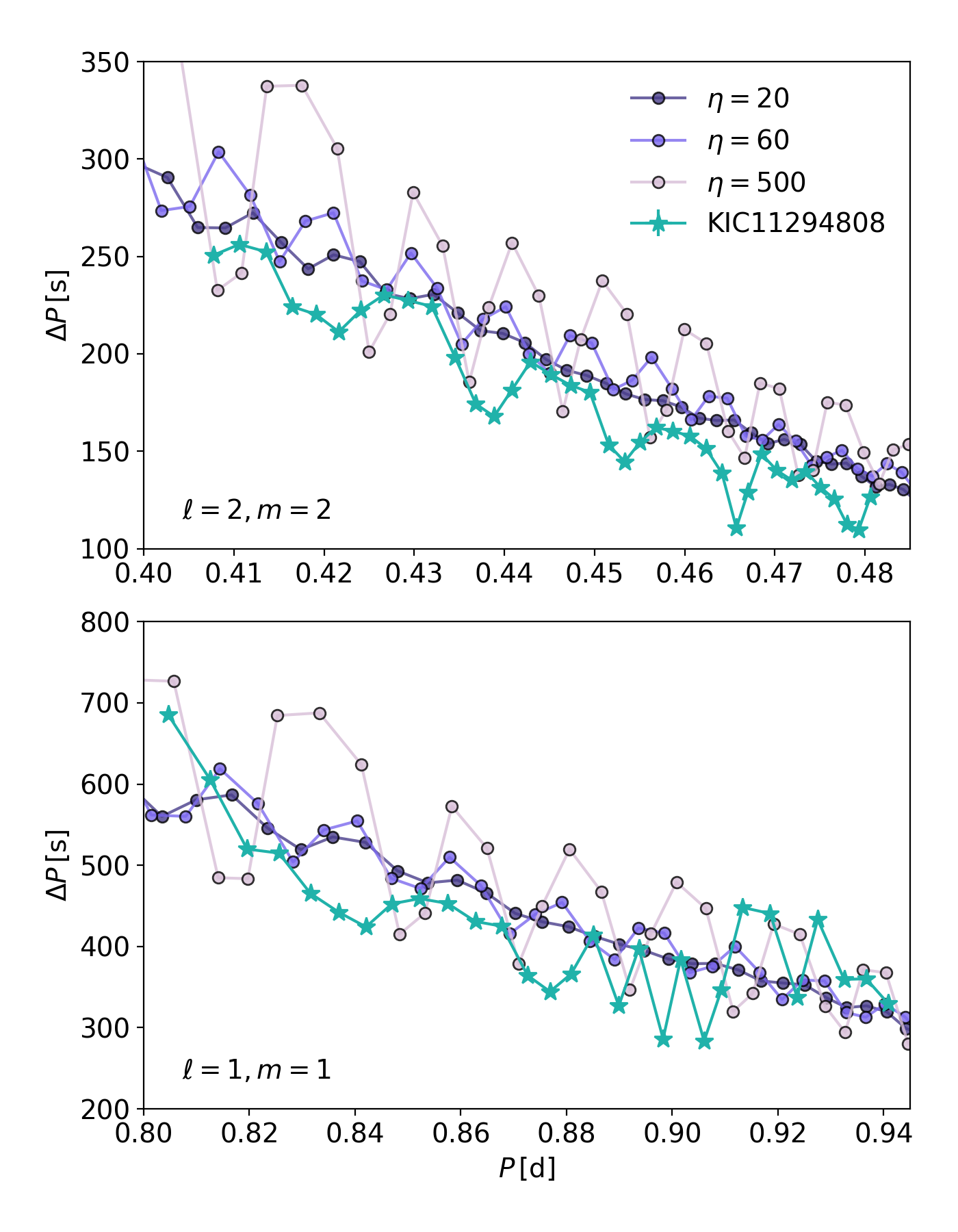}
    \caption{The predicted period spacing patterns for different values of the $\eta$ parameter in Eq.~\ref{eq:Dmix} for a model with a mass and $X_{\rm c}$ corresponding to the maximum likelihood estimate by \cite{Mombarg2021} in case of \kico. The observed period spacing patterns by \cite{VanReeth2015-spec} are shown by the sea green star symbols. The uncertainties of the observed values are typically smaller than the symbol size.   }
    \label{fig:PSP_kico}
\end{figure}

One particularly interesting star in the sample of \cite{VanReeth2015-spec} is \kico ($f_{\rm rot} = 0.77\,{\rm d^{-1}}, M_\star = 1.655\,{\rm M_\odot}, X_{\rm c} = 0.19$), for which period spacing patterns of dipole $(\ell = 1,m=1)$ modes and quadrupole $(\ell=2,m=2)$ modes were observed. The periods of the quadrupole modes exhibit clear recurring dips, indicative of trapped modes. Therefore, we use this star to find a reasonable value for $\eta$, that produces dips in the period spacing pattern for a model with the parameters found by \cite{Mombarg2021}. We use a scaling of $\alpha_{\rm ov} \approx 10f_{\rm ov}$ \citep{ClaretTorres2017}, as in this work, we model the CBM with convective penetration. On one hand, when $\eta$ is set too low ($\lesssim 20$, see Fig.~\ref{fig:PSP_kico}), helium settling is not counteracted efficiently, thereby stabilizing the chemical gradient near the core boundary. On the other hand, if $\eta$ is set too high (around 5000, \redits{not shown}), chemically homogeneous evolution occurs. As shown in Fig~\ref{fig:PSP_kico}, a value of $\eta =60$ provides a good estimate for \kico. We stress that the predicted locations of the dips in the period spacing pattern are extremely sensitive to the stellar mass, age, and input physics, and that the aim of this paper is not to precisely match the periods, but to investigate if models with shear mixing and radiative levitation can reproduce the global morphology of the observed period spacing pattern.  

\section{Oscillation spectra} \label{sec:psp}
\begin{figure}
    \centering
    \includegraphics[width = \columnwidth]{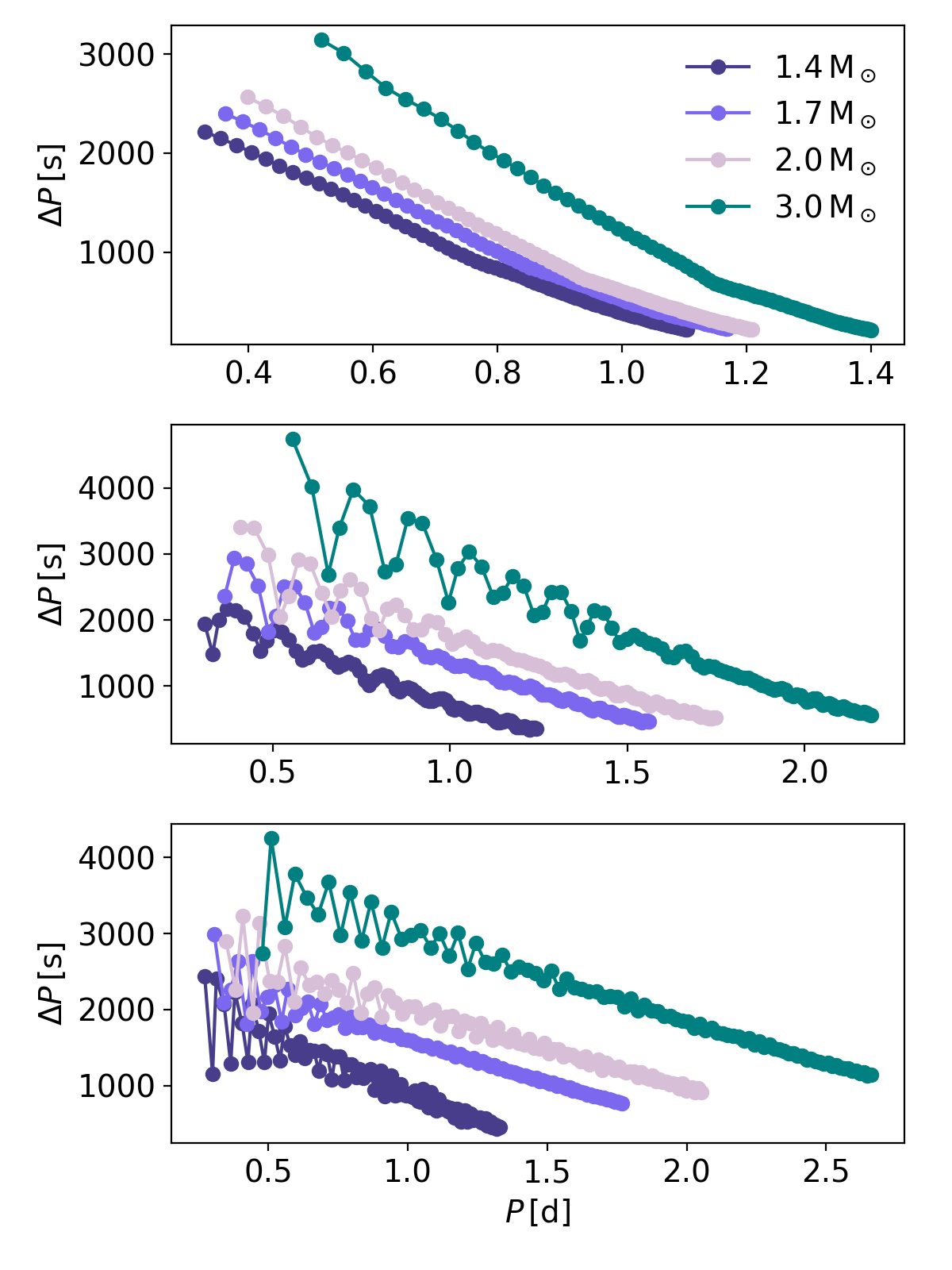}
    \caption{Period spacing patterns for models with shear mixing and atomic diffusion (including radiative levitation) at \xci = 0.95 (top panel), \xci = 0.45 (middle panel), and \xci = 0.10 (bottom panel). For all models $Z_{\rm ini} = 0.015$, $\alpha_{\rm ov} = 0.15$, and $\Omega = 0.2\Omega_{\rm bk}$. }
    \label{fig:PSP-Xc}
\end{figure}

We have computed \mesa models for masses of 1.4\Msun, 1.7\Msun, 2.0\Msun, and 3.0\Msun covering the entire \gDor mass regime and the lowest part of the SPB mass regime. For each model, we have computed the predicted periods at \xci = 0.95, 0.45, 0.10. Moreover, we have tested three different values for the initial metallicity, \zini, namely 0.010, 0.015, and 0.020. We refer to \cite{Mombarg2021} and \cite{Mombarg2020} for more details on our \mesa setup. One difference between the setup used in this work and the aforementioned ones, is that here we describe the CBM by convective penetration, following the results from \cite{Pedersen2021}. It should be noted, however, that no clear distinction between an exponential and step overshoot profile\footnote{The temperature gradient is taken to be the radiative one, as opposed to convective penetration.} \edits{has yet been} made for \gDor stars \citep{Mombarg2019, Mombarg2021}. Our focus is on the prograde dipole modes and radial orders $n \in [-100,-10]$, as these are most commonly observed mode geometry in \gDor and SPB pulsators \citep[e.g.][]{VanReeth2016, Li2020, Pedersen2021}. The mode periods were computed using the stellar pulsation code \gyre \citep[v5.2;][]{Townsend2003, Townsend2013}. In Fig.~\ref{fig:PSP-Xc}, we show the predicted period spacings, $\Delta P_{n} = P_{n+1} - P_{n}$, as a function of period, $P_{n}$, for the three different values of \xci. To produce mode trapping when atomic diffusion (including radiative levitation) is taken into account, a low mixing efficiency close to the core boundary is required, and a high efficiency further outward. Our models show this can be realized by the prescription given in Eq.~\ref{eq:Dmix}, as we observe mode trapping for all stellar masses studied here when we choose $\eta = 60$. The prescription for REM used in this work also \redits{requires} much higher diffusion coefficients outside the shear zone, solving the stark contrast between theoretical predictions and the very low diffusion coefficients that \redits{\cite{VanReeth2016} and} \cite{Mombarg2021} needed to explain the mode trapping in some stars in their sample, \edits{and the low values found by \cite{Moravveji2016}}, when the value of the diffusion constant was fixed throughout the star.

\begin{figure}
    \centering
    \includegraphics[width = \columnwidth]{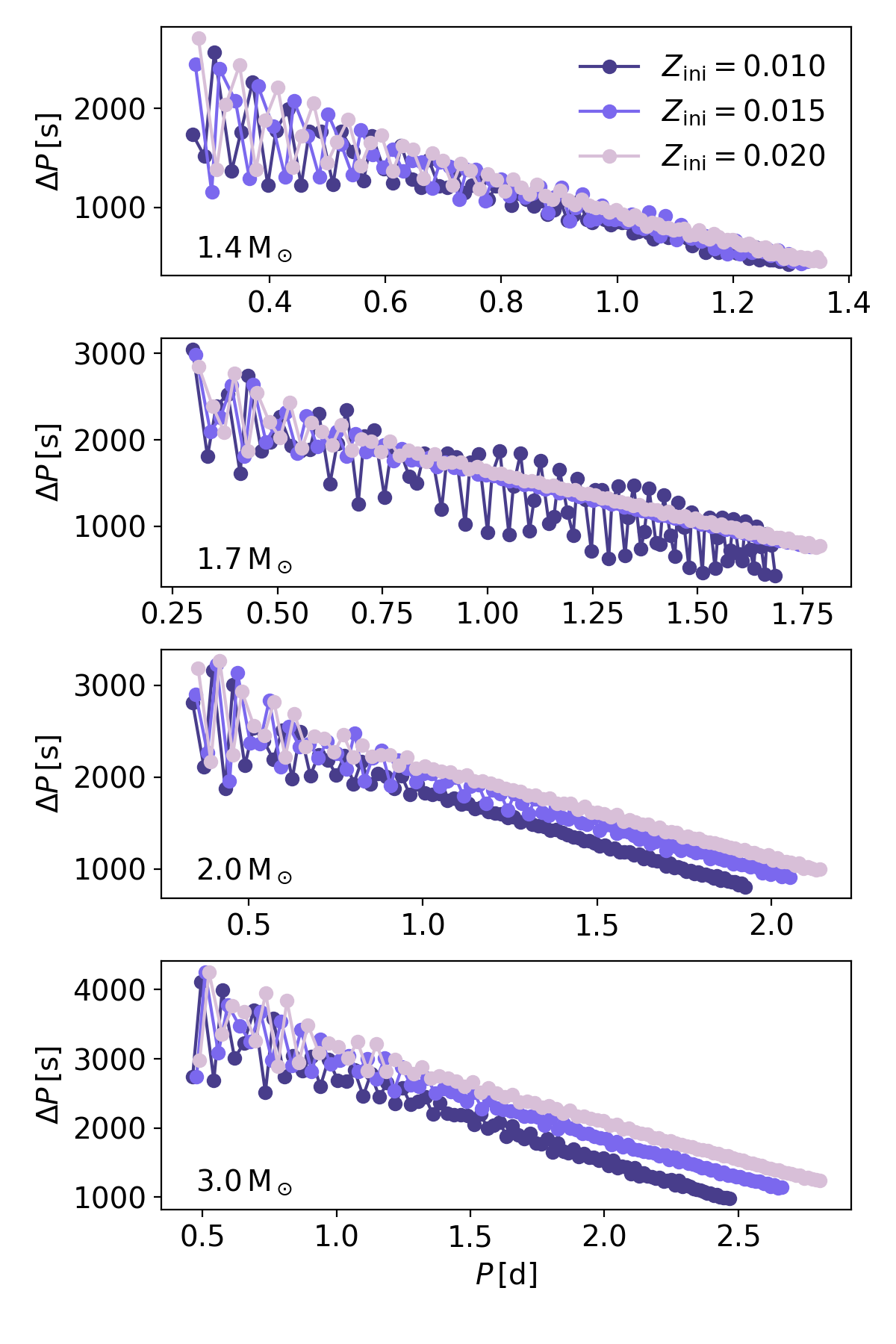}
    \caption{Same as Fig.~\ref{fig:PSP-Xc}, but for varying metallicity. For all models \xci = 0.10.}
    \label{fig:PSP-Z}
\end{figure}

In Fig.~\ref{fig:PSP-Z}, the effect of the initial metallicity is demonstrated for models with \xci = 0.10 (see Fig.~\ref{fig:PSP-ZXc0.45} in the appendix for models with \xci = 0.45). Since the initial helium abundance in our models is scaled with the metallicity, according to the chemical enrichment rate derived by \cite{Verma2019}, a higher \zini means that the helium settling near the core diminishes the chemical gradient to a larger extent, and as such, the dips in the period spacing pattern caused by mode trapping is less pronounced. Interestingly, we observe three sinusoidal modulations in the period spacings of the high-radial order modes of the evolved 1.7\dMsun model in the lowest metallicity case, whereas for higher metallicities, a nearly constant slope is seen. In the model with \zini = 0.010, additional dips in the REM profile are formed, which create several mode cavities in which modes can get trapped \edits{(see Fig.~\ref{fig:kernel} in the appendix)}. 

The signatures of mode trapping for models with \redits{combined} atomic diffusion and shear mixing are similar to those predicted by models with no (or very low) micro- and macroscopic REM, as either set of physics introduces an oscillatory behavior of the dips in the period spacing patterns, as shown by \cite{Miglio2008} and \cite{Bouabid2013}. Besides mode trapping, dips in the period spacing pattern can also be caused by coupling between gravito-inertial modes in the envelope and pure inertial modes in the convective core \redits{at specific mode frequencies} \citep{Ouazzani2020, Saio2021}. \newline

\begin{figure*}
    \centering
    \includegraphics[width = 0.95\textwidth]{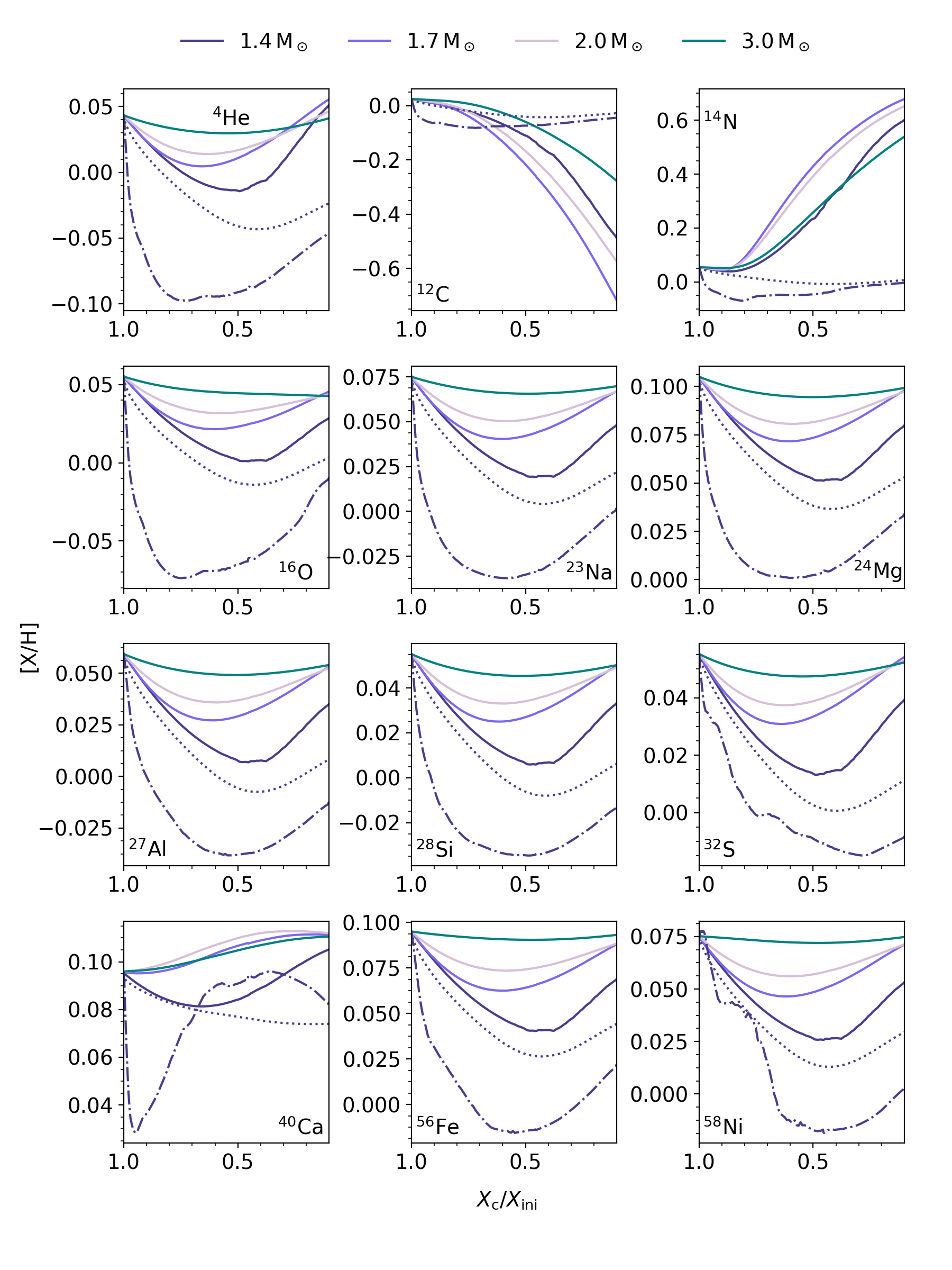}
    \caption{Predicted evolution of the surface abundances (scaled to the Sun) for models with \zini = 0.015 \redits{and $\eta = 60$}. The dotted lines correspond to a model with 1.4\dMsun and $\eta = 2$. The dashed-dotted lines for a 1.4\dMsun model without any macroscopic mixing ($\eta = 0$).}
    \label{fig:XH_evol}
\end{figure*}

\begin{figure}
    \centering
    \includegraphics[width = \columnwidth]{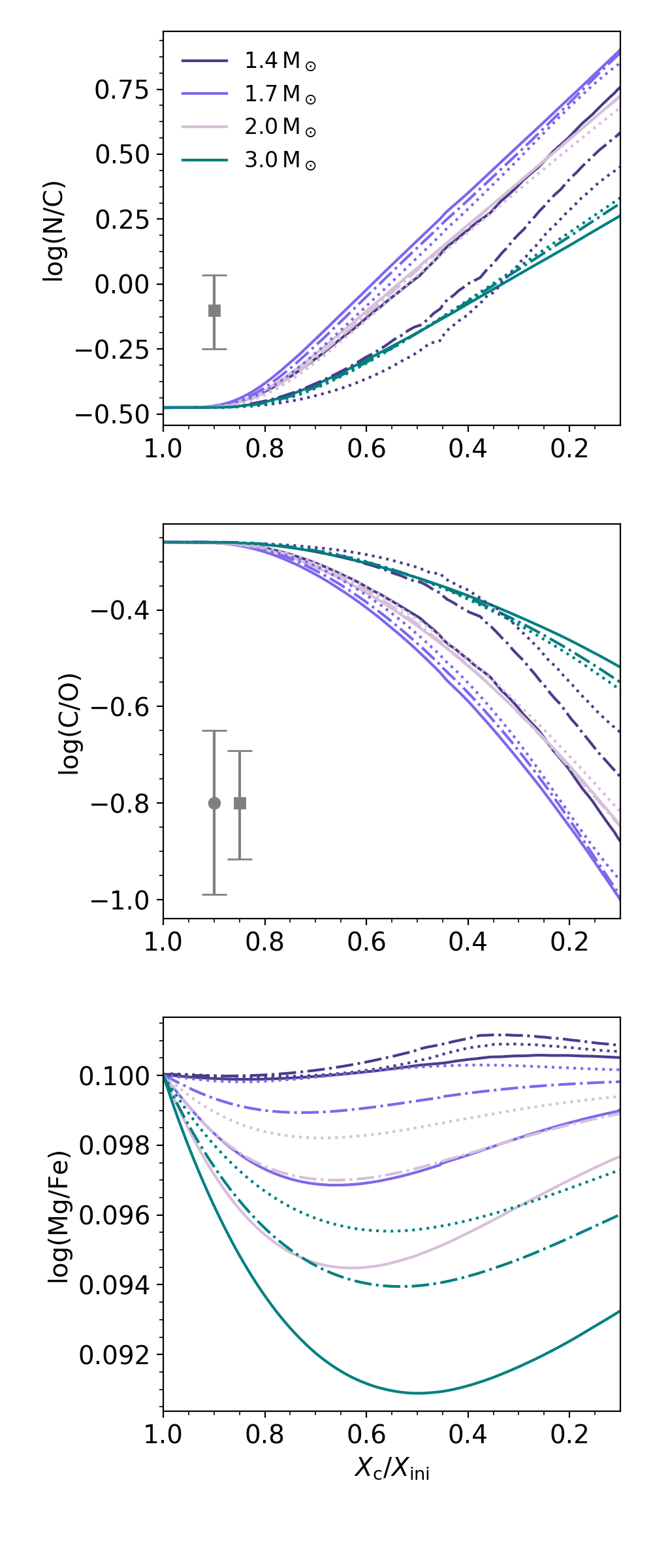}
    \caption{Predicted evolution of several surface abundance ratios for models with \zini = 0.010 (solid lines), 0.015 (dash-dotted lines), and 0.020 (dotted lines). The most precise measurements achieved by \citep{Gebruers2021} for SPB and \gDor stars are indicated in gray with a square and circle symbol, respectively. The most precise measurements of the Mg/Fe ratio are around 0.6\,dex, and are thus much larger than the predicted change of this ratio along the evolution.}
    \label{fig:abun_ratios}
\end{figure}

\section{Surface abundances} \label{sec:abun}
\edits{Additionally, the inclusion of \redits{REM} allows us to study the evolution of the surface abundances, which provide extra constraints on the global efficiency of the chemical mixing. } \cite{Dotter2017} already showed that atomic diffusion and radiative levitation affect the surface abundances throughout stellar evolution significantly for stars with masses between 0.5\Msun and 1.5\Msun. In Fig.~\ref{fig:XH_evol}, we show for our models the predicted surface abundances\footnote{Presented as ${\rm [X/H]} = \log(X_{\rm X}/X_{\rm H}) - \log(A_{\rm X}/A_{\rm H}) - \log\,\epsilon_{\rm X, \odot} + 12$, with $X_{\rm X}$ the mass fraction, and $A_{\rm X}$ the atomic mass. \redits{For the values of $\log\,\epsilon_{\rm X, \odot}$, the Solar composition from \citet[][$Z_\odot = 0.0139$]{Asplund2021} is used.}} as a function of age for elements He, C, N, O, Na, Mg, Al, Si, S, Ca, Fe, and Ni. The predicted evolution of the [Fe/H] abundance of the 1.4\dMsun model from \cite{Deal2020}, who also include radiative levitation and rotational mixing in their models, shows a similar trend compared to our findings. That is, the [Fe/H] abundance decreases during the first part of the MS and increases again during the second part of the MS. The same behavior is seen for elements heavier than oxygen, which do not partake in the CNO-cycle and thus have a constant bulk abundance. Around halfway through the MS, the radiative acceleration of iron peak elements exceeds the local gravitational acceleration, preventing these elements from sinking towards the core, resulting in an increase of the metal surface abundances. Yet, for the models with 1.7\Msun and up, the radiative levitation of calcium is strong relatively near the surface, and a constant increase in the surface abundance along the MS is observed. The fact that we \edits{see} changes in surface abundances throughout the evolution indicates that, even the high mixing efficiencies in the envelope used in this work are not enough to completely dominate over the effects of radiative levitation. Yet, the typical excursions in [X/H] are about a factor ten less than what was predicted in the absence of strong REM by \cite{Mombarg2020}. For the other two metallicity values \redits{(\zini =  0.010 and 0.020)}, we find similar behavior of the surface abundances. As typical precisions for [X/H] derived in F-type stars is at best on the order of 0.1\,dex \citep{Gebran2010,Niemczura2015,Takada-Hidai2017, Gebruers2021}, we expect that most \gDor stars will not show clear signs of heavy element depletion at the surface, as most have moderate to fast rotation frequencies \citep{VanReeth2016, VanReeth2018}. Indeed, spectroscopic studies of \gDor and SPB stars show most of them have surface compositions similar to the Sun \citep{Kahraman2016, Gebruers2021}. 

\redits{Aside from microscopic mixing, efficient REM due to rotational} mixing can transport the CNO elements from the core towards the surface. Hydrogen burning via the CNO-cycle occurs through a chain of several reactions, where the conversion of nitrogen to oxygen is slow and causes in buildup of nitrogen and a depletion of carbon, while the amount of oxygen stays roughly the same. We predict that this change can be observed as the gravitational settling of these elements is not efficient enough \edits{to cause depletion of the CNO elements at the surface.} In Fig.~\ref{fig:abun_ratios}, the ratios of N/C and C/O are shown, as well as Mg/Fe and O/Fe. Observational measurements of log(N/C) or log(C/O) could serve as additional constraints on the stellar age, if determined with precisions of $\sim$0.2\,dex and 0.1\,dex, respectively. The ratio Mg/Fe is typically the abundances ratio which can be determined most precisely in F-type stars. Our models predict this ratio remains almost constant throughout the MS, which is consistent with ratios inferred by \cite{Gebruers2021} for a sample of 91 \gDor stars. 

Additionally, the predicted evolution of the surface abundances for a 1.4\dMsun model with $\eta = 2$ (i.e. the horizontal diffusion coefficient is of the same order as the thermal diffusivity) is shown in Fig.~\ref{fig:XH_evol} (dotted lines). We find that the excursions in [X/H] of O, Na, Mg, Al, Si, S, Fe, and Ni are not significantly increased, compared to typical precisions that can be achieved with spectroscopy. The lower mixing efficiency at the outer part of the envelope can no longer drive Ca to the surface around mid-MS, resulting in Ca depletion when the star evolves. The diffusion coefficient just below the surface convection zone of the model with $\eta = 2$ is about two orders lower than that of the model with $\eta = 60$ (i.e. the thermal diffusivity is \edits{smaller} than the horizontal diffusion). For C and N, we find that the mixing becomes too low to induce any changes in the abundances at the surface. In more evolved stars, the C/H ratio could be used to put a lower limit on the mixing just outside of the CBM zone, when the bulk metallicity is well-known and the initial composition is assumed to be close to the solar composition. For models without shear mixing, the maximum depletion of the heavier elements is stronger, and underabundances in Mg and Fe (Mg/Fe ratio remains stable) could be observed in mid-MS stars when precisions of $\sigma_{\rm [X/H]} \leq 0.1\,{\rm dex}$ can be achieved. \redits{The 1.4\dMsun model without macroscopic mixing (dash-dotted lines in Fig.~\ref{fig:XH_evol}) does not show complete helium depletion, even though the radiative levitation for this element is extremely weak. The settling of helium at the surface of this model is impeded by the formation of a surface convection zone \citep[cf.][]{VermaSilva2019}, as is shown in Fig.~\ref{fig:helium_profiles} in the appendix.  }

\section{Numerical uncertainties} \label{sec:numerical}
As the profile of the Brunt-V\"ais\"al\"a frequency is used in the computation of the mixing profile, the evolution becomes sensitive to the mesh resolution of the stellar model. For simple models without radiative levitation or shear mixing, the difference in the predicted mode periods at some point becomes smaller than the typical observational uncertainty \edits{on the period} when more cells are added to the model. However, when radiative levitation is taken into account, the mesh resolution is limited, since each cell has its own composition, which makes it significantly more difficult for the solver to converge to a model within the tolerances. We computed the best model of \kico with three different mesh grids with 2845 (M1), 3010 (M2), and 3109 (M3) cells\footnote{\mesa adapts the total number of cells at each time step. The numbers listed are specifically for the mentioned models.}, where the extra cells are placed around the large change in chemical composition just outside the core. The differences in the predicted mode periods by model M2 and M3 are roughly the same as the period differences between model M1 and M2, as shown in Fig.~\ref{fig:resol} in the appendix, indicating our predictions of the mode periods are limited to a numerical uncertainty of $\sim$60~sec on the low-radial order \edits{(shorter period)} modes, and $\sim$30~sec for the high-radial order \edits{(longer period)} modes. \redits{In Fig.~\ref{fig:mode_trapping}, we show rotational kernels overplotted on the g-mode cavity to illustrate that these the dips in the period spacing patterns are indeed caused by trapped modes, and are not the result of theoretical uncertainties. We emphasize that these uncertainties are caused by the choice of the meshing, and that it is not the noise on the individual periods computed from a single equilibrium model.} \edits{Even though these numerical uncertainties are, compared to the mode periods, smaller than 0.2\%, they are still several times larger than the extremely small uncertainties typically achieved with the {\it Kepler} nominal mission data.  } 

\section{Discussion and conclusion} \label{sec:discussion}
This study demonstrates the added value of \redits{integrating a} 1D \edits{stellar evolution code} and a 2D \edits{stellar structure} code to provide a more accurate description of \redits{the evolution of} chemical mixing in the envelope of rotating stars with a convective core. The predictions presented in this work serve as a valuable guide for more detailed calibration of the envelope mixing in intermediate-mass gravity-mode pulsators, as \cite{Pedersen2021} already showed the need of varying diffusion coefficients throughout the envelope in order to explain the observed pulsations of a sample of 26 SPB stars. \redits{The treatment of chemical mixing presented in this work allows for mode trapping, even when microscopic diffusion in the form of atomic diffusion (including radiative levitation) is taken into account. Moreover, it results in realistic and much larger diffusion coefficients in the outer envelope, compared to when microscopic diffusion and shear mixing are ignored.} The inclusion of radiative levitation (and consistent Rosseland opacity) makes the morphology of the period spacing pattern more dependent on the bulk metallicity. When shear mixing is the dominant source of macroscopic mixing, the iron abundance is expected to be relatively stable over the main-sequence evolution, and thus can be used as a tracer for the bulk metallicity of the star. We predict that in some cases, stars with subsolar metallicity (${\rm [M/H]} \textless -0.15$) show a very characteristic modulation of the period spacings in the asymptotic regime. The estimated mass and age of the star KIC\,7380501 by \cite{Mombarg2021} are close to that of the 1.7\dMsun model in Fig.~\ref{fig:PSP-Z}, and its metallicity is subsolar (${\rm [M/H]} = -0.21 \pm 0.06$; \citealt[][]{Gebruers2021}). The morphology of \redits{the period spacing pattern of} KIC\,7380501 is particularly complex to model \edits{(see Fig.~\ref{fig:psp_kic738} in the appendix)}, where some parts of the period spacing pattern, consisting of three modes, show a much larger slope compared the general slope of the pattern. This could be explained by the pattern predicted \redits{from the aforementioned equilibrium models}, if only a few consecutive modes are observed, and so radial orders are suppressed. 

Besides mixing induced by rotation, it has been shown that internal gravity waves (IGWs) generated \edits{by turbulent core convection} can also efficiently transport material throughout the radiative envelope \citep{LecoanetQuataert2013, RogersMcElwaine2017}. The work of \cite{RogersMcElwaine2017} predicts the efficiency of the mixing induced by IGWs to scale as $D_{\rm IGW}(r) \propto \rho(r)^{-\zeta}$, where $\zeta$ is somewhere between $\frac{1}{2}$ and 1. Hence, we expect that for stars with trapped modes, IGW mixing is not dominant, as the current simulations do not predict a large variation of the mixing coefficient near the core. Yet, for stars that show no signs of mode trapping in their oscillation spectra, mixing induced by IGWs might be the dominant source. A density dependent scaling of the diffusion coefficient (that is constant in time) is hard to distinguish from a constant diffusion coefficient, based on the periods of gravity modes \citep{Pedersen2018}. In that case, the surface abundances may hold more discriminating power when microscopic mixing is treated consistently, as done in this work. The effect of IGW mixing on the mode periods and surface abundances will be addressed in a future paper. 

The NASA TESS mission \citep{Ricker2015} covers a much larger part of the sky compared to {\it Kepler} \citep{Borucki2010}, and therefore provides a unique opportunity to study stars in different metallicity regimes. Period spacing patterns in 128 \gDor stars in the TESS Southern Continuous Viewing Zone have been extracted by Garcia et al. (submitted). These stars are overall brighter than those observed by {\it Kepler}, making them suitable for high-resolution spectroscopic follow-up studies.   
\clearpage

\bibliography{main}{}
\bibliographystyle{aasjournal}

\begin{acknowledgments}

The research leading to these results has received funding from the European
Research Council (ERC) under the European Union’s Horizon 2020 research and
innovation programme (grant agreement N$^\circ$670519: MAMSIE) and from the
KU\,Leuven Research Council (grant C16/18/005: PARADISE).
The computational resources and services used in this work were provided by the VSC (Flemish Supercomputer Center), funded by the Research Foundation - Flanders (FWO) and the Flemish Government – department EWI. JSGM, TVR and MM gratefully acknowledge support from the Research Foundation Flanders (FWO) through grants V429020N, 12ZB620N and 11F7120N, respectively. MR acknowledge the support of the French Agence Nationale de la Recherche (ANR), under grant ESRR (ANR-16-CE31-0007-01). Part of the numerical computations have been performed using the HPC resources from CALMIP (grant 2020-P0107), which is gratefully acknowledged. We thank Dominic M. Bowman for his comments on the manuscript. A special thanks to Bill Paxton and his \mesa team, and also to Rich Townsend and his \gyre team for all the efforts put into these projects that have been paramount to this work. \redits{We are grateful for the fast response of the anonymous referee, and for their comments which have helped to improve the clarity of this manuscript.}

\software{\mesa~\citep[r11701;][]{Paxton2011, Paxton2013, Paxton2015, Paxton2018, Paxton2019},~\texttt{MESASDK}~\citep{Townsend2019-linux,Townsend2019-mac},~\gyre~\citep[v5.2;][]{Townsend2013, Townsend2018},~\texttt{matplotlib}~\citep{Hunter2007},~\texttt{numpy}~\citep{Harris2020}}
\end{acknowledgments}

\appendix 
\renewcommand\thefigure{\thesection.\arabic{figure}} 
\section{Benchmark results}
In Fig.~\ref{fig:grad}, we show a benchmark test of our novel implementation for computing radiative accelerations and consistent Rossland mean opacities, compared to the implementation in \mesa r11701.
\setcounter{figure}{0} 
\begin{figure}[htb]
    \centering
    \includegraphics[width = \columnwidth]{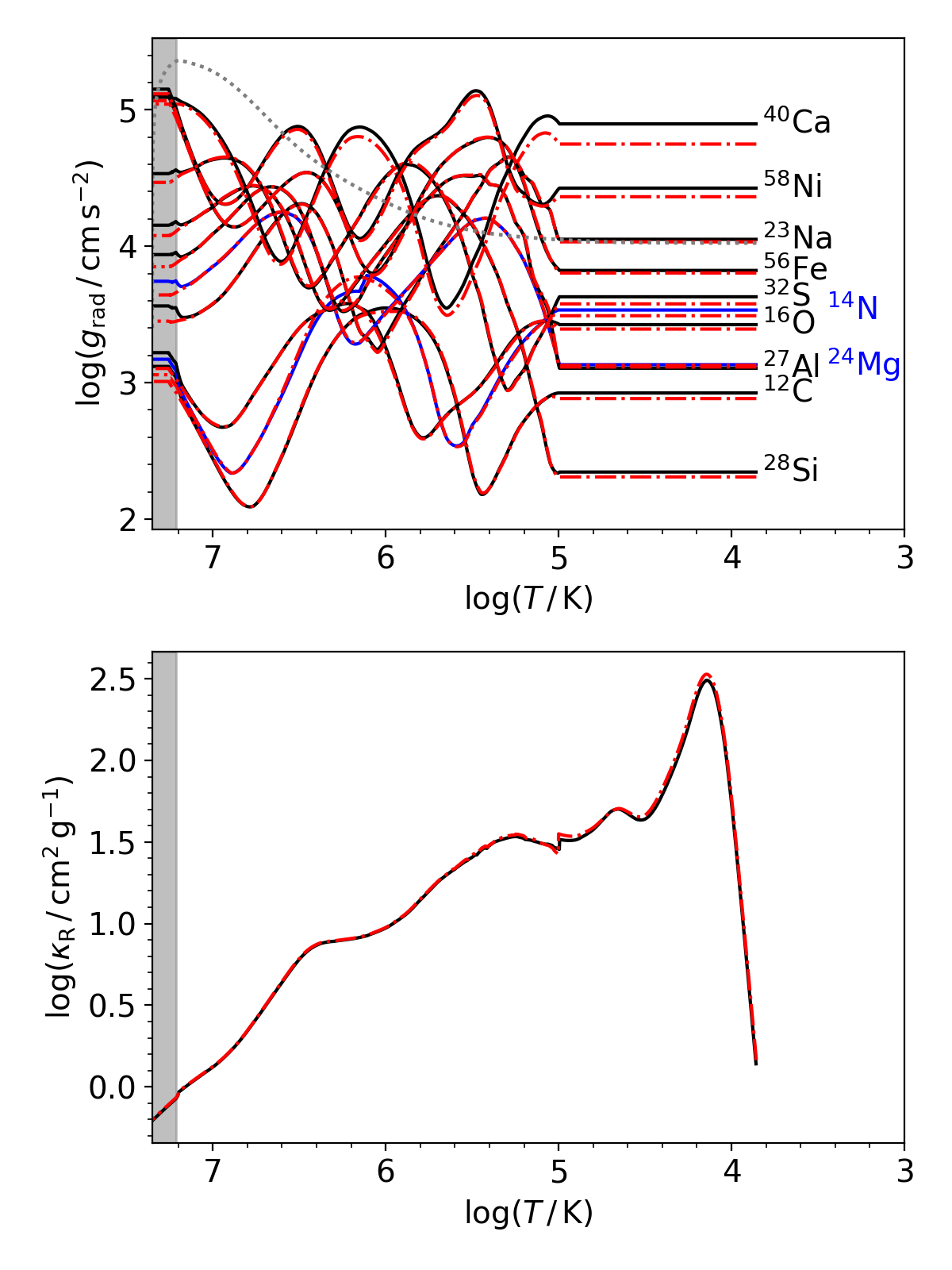}
    \caption{A comparison between the radiative accelerations (top panel) and Rosseland mean opacity (bottom panel) computed with the method presented in this paper (solid lines) and the routines from \citep{Hu2011} implemented in \mesa r11701 (dash-dotted lines). The computations are based on a 1.7\dMsun model at $X_{\rm c} = 0.325$ (\zini = 0.015). The extent of the convective core is highlighted by the shaded region. The local gravitational acceleration is indicated by the gray dotted line. }
    \label{fig:grad}
\end{figure}

\section{Predicted period spacings mid-MS}
\setcounter{figure}{0} 

\begin{figure}
    \centering
    \includegraphics[width = \columnwidth]{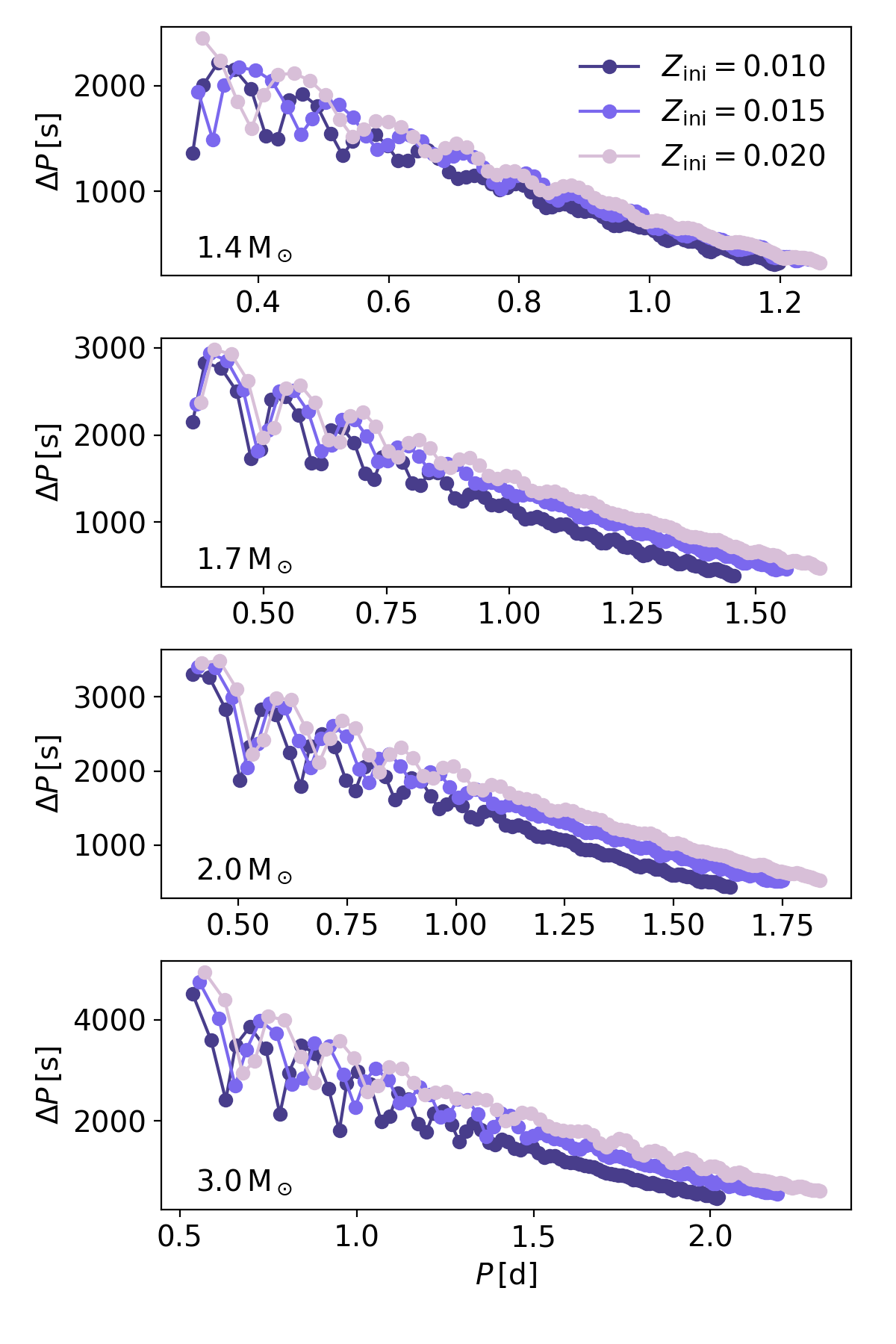}
    \caption{Predicted period spacing patterns for models with shear mixing and radiative levitation, for different metallicities. For all models \xci = 0.45.}
    \label{fig:PSP-ZXc0.45}
\end{figure}

In Fig.~\ref{fig:PSP-ZXc0.45}, we show the effect of the initial metallicity on the predicted period spacing patterns for models of mass 1.4\Msun, 1.7\Msun, 2.0\Msun, and 3.0\Msun, roughly halfway through the MS.

\section{Rotational kernels}
\setcounter{figure}{0} 
Fig.~\ref{fig:kernel} shows the predicted mixing profile, and overplotted the rotational kernel of a g-mode with $(\ell, m, n_{\rm pg}) = (1,1,-75)$. For the model with \zini = 0.010 (top panel), this mode is trapped, whereas for the model with \zini = 0.015 mode trapping is not observed.
\begin{figure}
    \centering
    \includegraphics[width = \columnwidth]{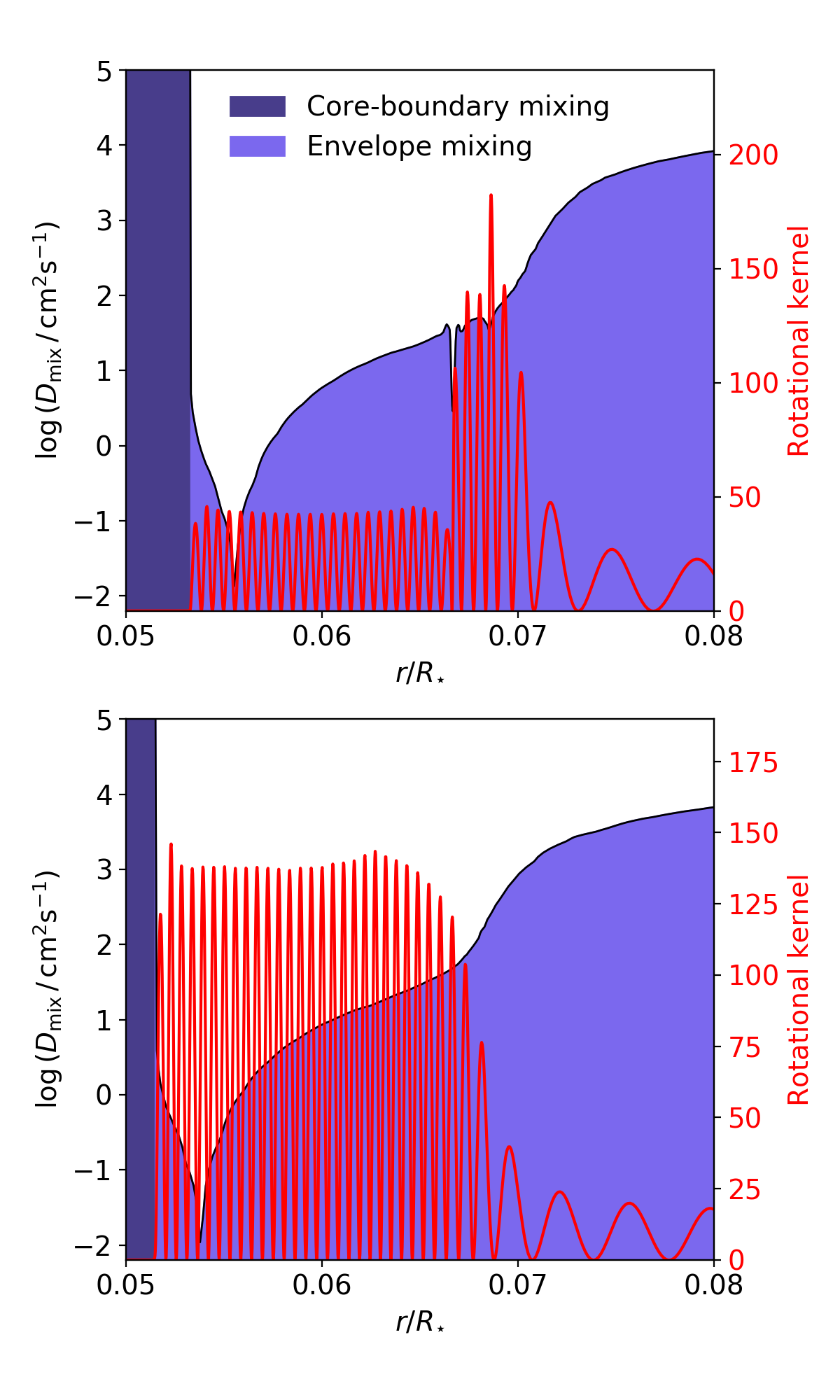}
    \caption{Mixing profiles and unnormalized rotational kernels for radial order $n_{\rm pg} = -75$  \cite[Eq. 3.356 in][]{Aerts2010} of the 1.7\dMsun models shown in the second panel from the top in Fig.~\ref{fig:PSP-Z} for $Z_{\rm ini} = 0.010$ (left panel) and $Z_{\rm ini} = 0.015$ (right panel).}
    \label{fig:kernel}
\end{figure}

\section{Chemical profiles of helium}
\setcounter{figure}{0} 

\begin{figure}
    \centering
    \includegraphics[width = \columnwidth]{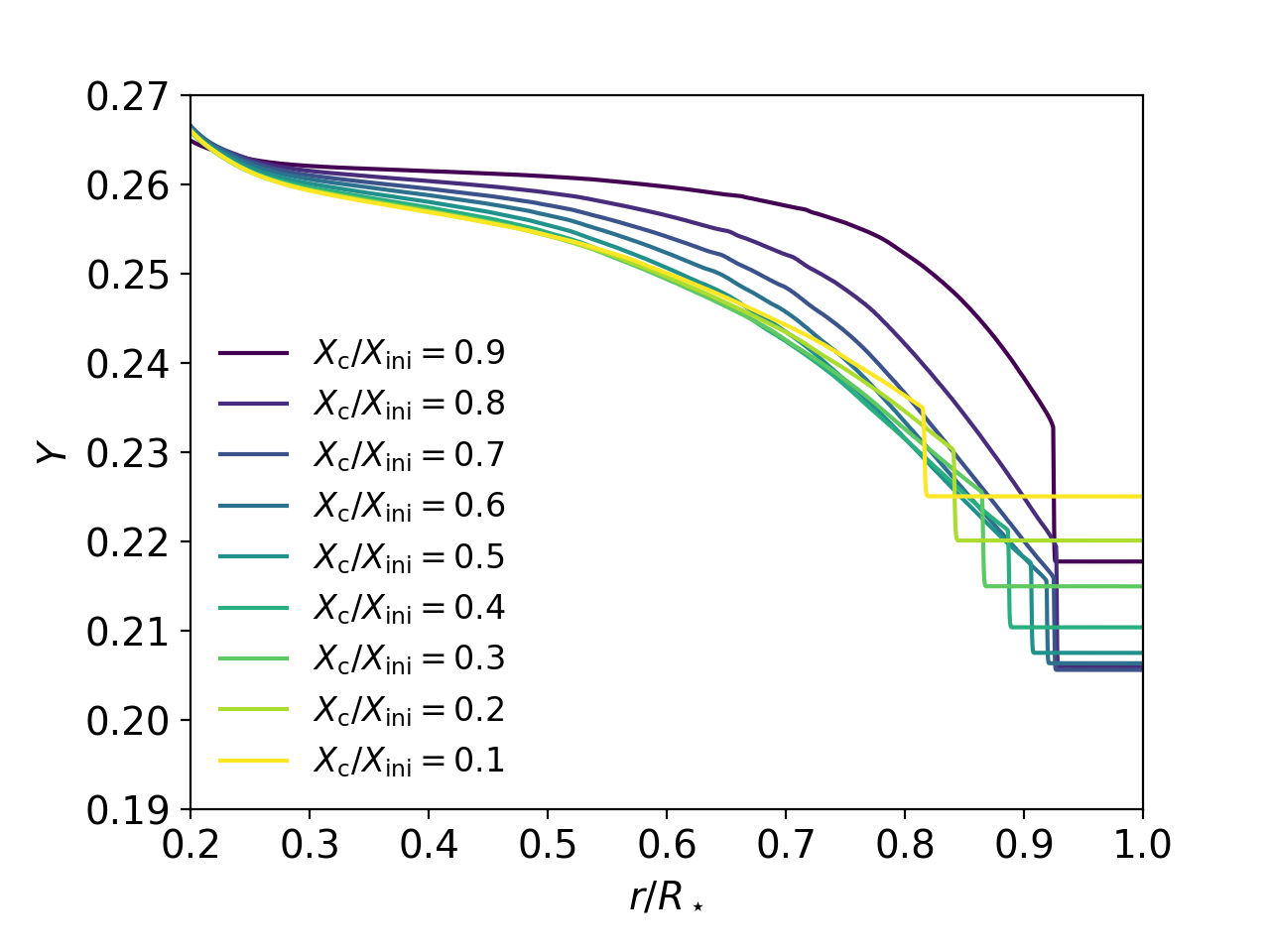}
    \caption{\redits{The helium mass fraction ($Y$) profiles of a 1.4\dMsun model (\zini = 0.015) for different times during the MS evolution. The outer part where the helium mass fraction is constant indicates the surface convection zone. }  }
    \label{fig:helium_profiles}
\end{figure}

\redits{In Fig.~\ref{fig:helium_profiles}, we show the helium mass fraction throughout the star at different times during the MS for a 1.4\dMsun model without macroscopic mixing (i.e. only microscopic mixing, dash-dotted lines in Fig.~\ref{fig:XH_evol}).}

\section{Theoretical uncertainty}
\setcounter{figure}{0} 
Fig.~\ref{fig:resol} demonstrates how the predicted mode periods (top and middle panel) and the Brunt-V\"ais\"al\"a frequency (bottom panel) are influenced by the number of cells of the \mesa equilibrium model. \redits{In Fig.~\ref{fig:mode_trapping}, we show in the corresponding rotational kernels of the shortest-period modes of the model with 2845 cells that is shown in Fig.~\ref{fig:resol}.}  
\begin{figure}
    \centering
    \includegraphics[width = \columnwidth]{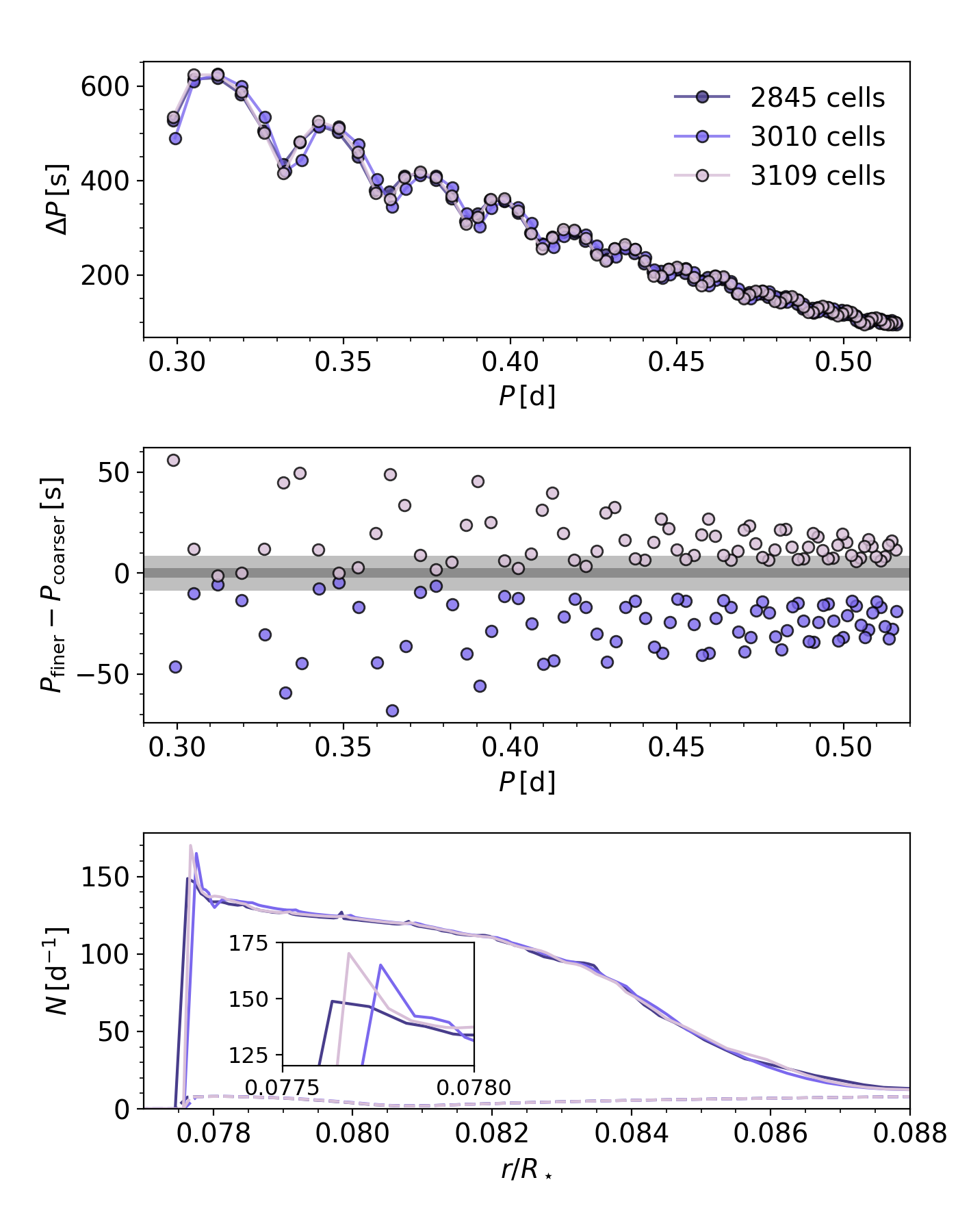}
    \caption{The effect of the number of cells used in the model, with 1.655\Msun, \zini = 0.02, \aov = 0.05, and $X_{\rm c} = 0.356$ on the predicted period spacing pattern (top panel, $(\ell = 2, m = 2)$). The middle panel shows the difference in period per radial order where the color corresponds to the model with a finer meshing. The light shaded region indicates the typical uncertainty on the period across the sample of \cite{VanReeth2015-spec}, the darker shaded area the average uncertainty for \kico. \redits{The bottom panel shows the corresponding total Brunt-V\"ais\"al\"a frequency (solid lines) and its structure component ($N_{\rm struc}$, dashed lines).}}
    \label{fig:resol}
\end{figure}

\begin{figure}
    \centering
    \includegraphics[width = \columnwidth]{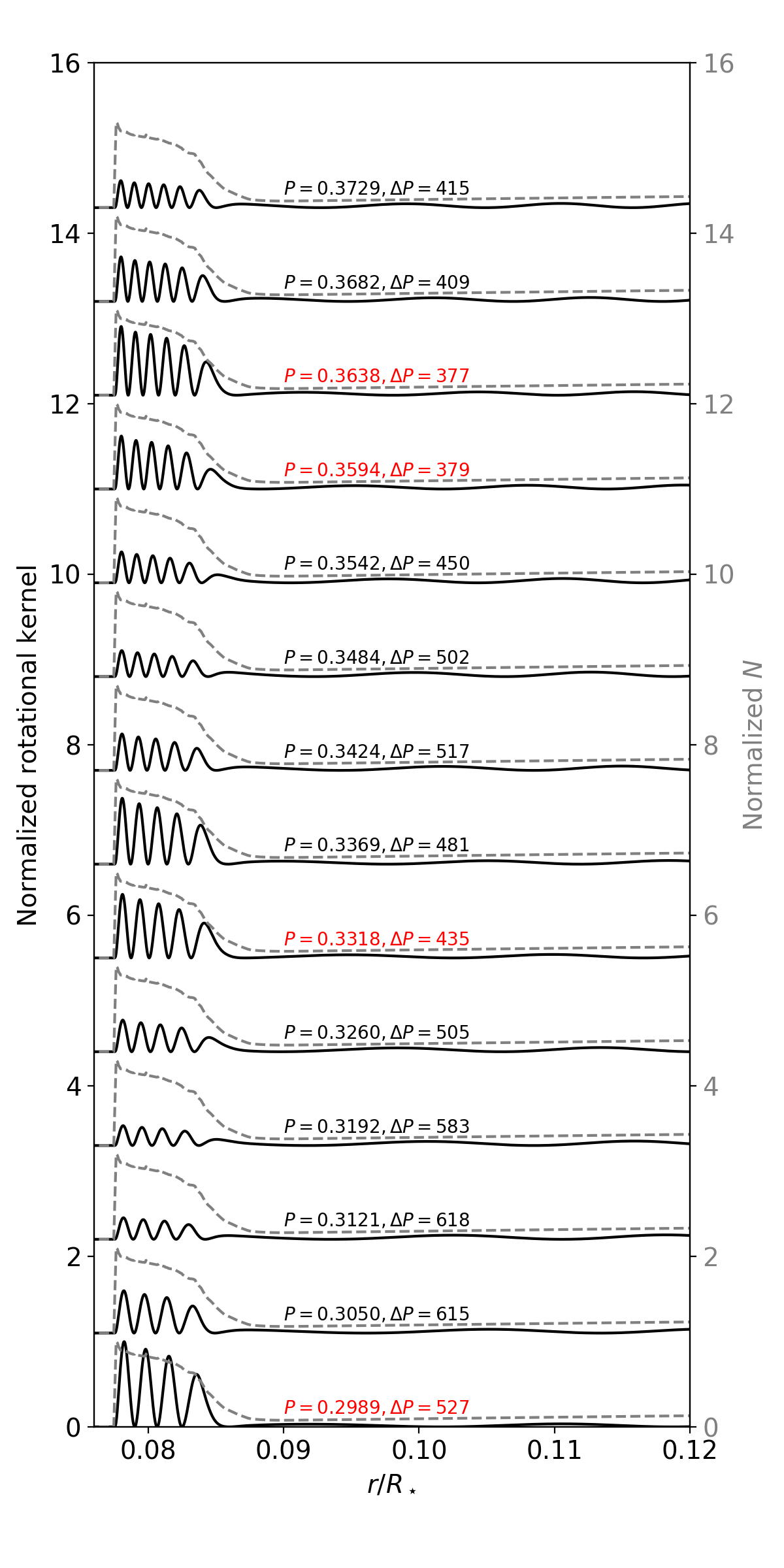}
    \caption{\redits{Rotational kernels normalized to the largest maximum amplitude of the kernels for the shortest-period modes of the model with 2845 cells. The corresponding mode period (in days) and period spacing (in seconds) are also shown. The modes which correspond to the dips in the period spacing pattern shown in the top panel of Fig.~\ref{fig:resol} are highlighted in red. The gray dashed lines indicate the Brunt-V\"ais\"al\"a frequency, normalized to one.}}
    \label{fig:mode_trapping}
\end{figure}

\section{KIC\,7380501}
\setcounter{figure}{0} 
In Fig.~\ref{fig:psp_kic738}, we show the $(\ell, m) = (1,1)$ period pattern of KIC\,7380501 that was extracted by \cite{VanReeth2015-spec}. The modulations observed in this period spacing pattern are particularly ill-reproduced by the best model found by \cite{Mombarg2021}.
\begin{figure}
    \centering
    \includegraphics[width = \columnwidth]{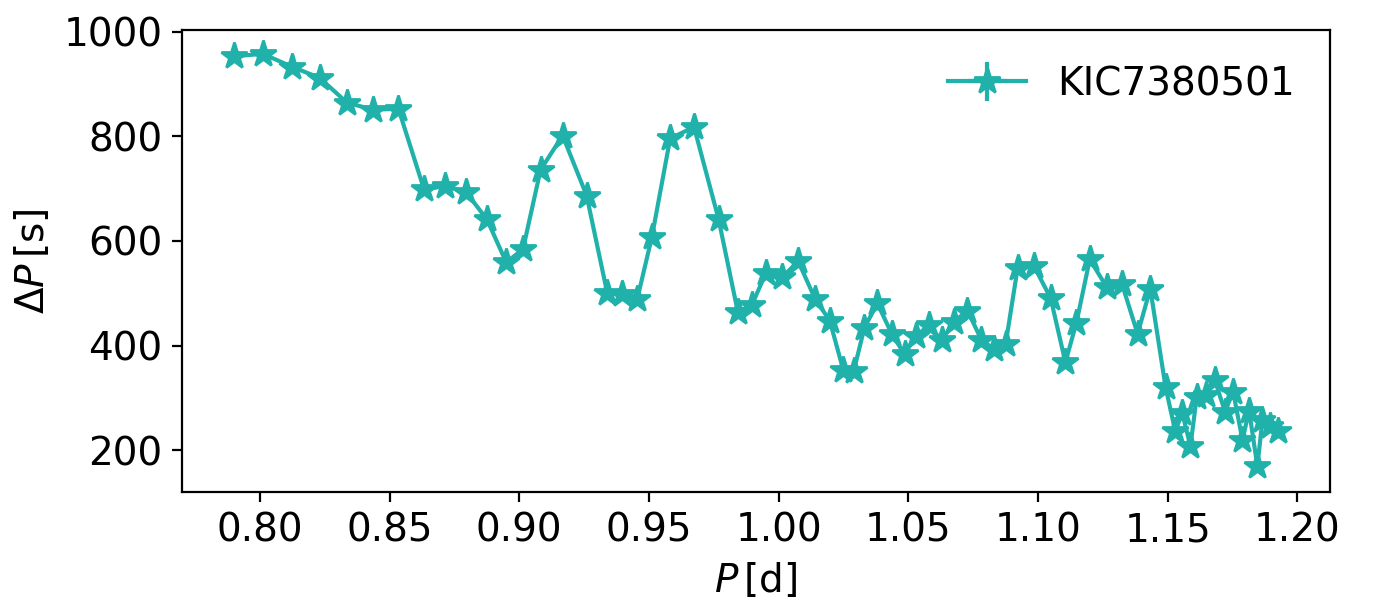}
    \caption{The observed period spacing pattern of KIC\,7380501 found by \cite{VanReeth2015-spec}. The uncertainties of the observed values are typically smaller than the symbol size. }
    \label{fig:psp_kic738}
\end{figure}


\end{document}